\documentclass[11pt,a4paper]{article}

\usepackage[english]{babel}

\pdfoutput=1 % if your are submitting a pdflatex (i.e. if you have
\RequirePackage{ifpdf}
\usepackage{amsmath}
\usepackage{mathtools}

\usepackage{jheppub}
\usepackage{pstricks}
\usepackage[final]{pdfpages}
\usepackage{ifpdf}
\usepackage{slashed}

\usepackage{hyperref}

\usepackage{color}
\usepackage{graphics}

\usepackage{etoolbox} % Load before 'breqn' package to make 'lineno' work
\usepackage{fixmath}

\usepackage{notoccite} % If you have \cite commands in \section-like commands, or in \caption, the
\usepackage{caption}
\usepackage{subcaption}
\usepackage{amsfonts}

\usepackage[normalem]{ulem}

\usepackage[compat=1.1.0]{tikz-feynman}
\usetikzlibrary{positioning,arrows}
\usetikzlibrary{decorations.pathmorphing}
\usetikzlibrary{decorations.markings}
\usetikzlibrary{shapes,shadows}

\usepackage{xcolor}

\definecolor{mygreen}{rgb}{0.0, 0.5, 0.0}
\definecolor{myorange}{rgb}{1.0, 0.49, 0.0}

\title{Two-loop mixed QCD-EW corrections to neutral current Drell-Yan}

\author[a]{Tommaso Armadillo,}
\author[b]{Roberto Bonciani,}
\author[a,c]{Simone Devoto,}
\author[a,c,d]{Narayan Rana,}
\author[a,c]{Alessandro Vicini}

\affiliation[a]{Dipartimento di Fisica ``Aldo Pontremoli'',
  University of Milano, I-20133 Milano, Italy}
\affiliation[b]{Dipartimento di Fisica, Universit\`a di Roma ``La Sapienza'' and INFN, Sezione di Roma, I-00185 Roma, Italy}
\affiliation[c]{INFN, Sezione di Milano, I-20133 Milano, Italy}
\affiliation[d]{Department of Physics, Indian Institute of Technology Kanpur, 208016 Kanpur, India}

\emailAdd{tommaso.armadillo@studenti.unimi.it}
\emailAdd{roberto.bonciani@roma1.infn.it}
\emailAdd{simone.devoto@unimi.it}
\emailAdd{narayan.rana@unimi.it}
\emailAdd{alessandro.vicini@mi.infn.it}

\abstract{
  We present the two-loop mixed strong-electroweak virtual corrections
  to the neutral current Drell-Yan process
 and we provide, in ancillary files, the explicit formulae of
 the infrared-subtracted finite remainder.
 The final state collinear singularities are regularised by the lepton mass.
 The evaluation of all the relevant Feynman integrals,
 including those with up to two internal massive lines,
 has been worked out relying on analytical and semi-analytical techniques,
 in the case of complex-valued masses.
}

\preprint{TIF-UNIMI-2022-1}

\begin{document}
\allowdisplaybreaks[4]
\unitlength1cm
\maketitle
\flushbottom

%************
% Definition
%************

\def\D{{\cal D}}
\def\DD{\overline{\cal D}}
\def\g{\overline{\cal G}}
\def\gm{\gamma}
\def\M{{\cal M}}
\def\ep{\epsilon}
\def\epm1{\frac{1}{\epsilon}}
\def\epm2{\frac{1}{\epsilon^{2}}}
\def\epm3{\frac{1}{\epsilon^{3}}}
\def\epm4{\frac{1}{\epsilon^{4}}}
\def\unM{\hat{\cal M}}
\def\ashat{\hat{a}_{s}}
\def\asmur{a_{s}^{2}(\mu_{R}^{2})}
\def\sigbar{{{\overline {\sigma}}}\left(a_{s}(\mu_{R}^{2}), L\left(\mu_{R}^{2}, m_{H}^{2}\right)\right)}
\def\sigbarn{{{{\overline \sigma}}_{n}\left(a_{s}(\mu_{R}^{2}) L\left(\mu_{R}^{2}, m_{H}^{2}\right)\right)}}
\def\unas{ \left( \frac{\hat{a}_s}{\mu_0^{\epsilon}} S_{\epsilon} \right) }
\def\rnM{{\cal M}}
\def\bt{\beta}
\def\cD{{\cal D}}
\def\cC{{\cal C}}
\def\ca{\text{\tiny C}_\text{\tiny A}}
\def\cf{\text{\tiny C}_\text{\tiny F}}
\def\ct{{\red []}}
\def\sv{\text{SV}}
\def\murOmu{\left( \frac{\mu_{R}^{2}}{\mu^{2}} \right)}
\def\bb{b{\bar{b}}}
\def\bt0{\beta_{0}}
\def\bt1{\beta_{1}}
\def\bt2{\beta_{2}}
\def\bt3{\beta_{3}}
\def\gm0{\gamma_{0}}
\def\gm1{\gamma_{1}}
\def\gm2{\gamma_{2}}
\def\gm3{\gamma_{3}}
\def\nn{\nonumber}
\def\l{\left}
\def\r{\right}
\def\nn{\nonumber \\&}

\def\asr{\left( \frac{\alpha_s}{4 \pi} \right)}
\def\asrhat{\left( \frac{\hat\alpha_s}{4 \pi} \right)}
\def\aem{\left( \frac{\alpha}{4 \pi} \right)}
\def\smu{\left( \frac{s}{\mu^2} \right)}
\def\J{{\cal J}}
\def\S{{\cal S}}
\def\I{{\cal I}}

\newcommand\as{\alpha_{s}}

\newcommand{\be}{\begin{equation}}
\newcommand{\ee}{\end{equation}}
\newcommand{\bea}{\begin{eqnarray}}
\newcommand{\eea}{\end{eqnarray}}
\newcommand{\smallw}{{\scriptscriptstyle W}}
\newcommand{\mt}{m_t}
\newcommand{\ml}{m_\ell}
\newcommand{\mw}{\mu_\smallw}
\newcommand{\mwsq}{\mu_\smallw^2}
\newcommand{\mwc}{\mu_{\smallw 0}}
\newcommand{\smallz}{{\scriptscriptstyle Z}}
\newcommand{\mz}{\mu_\smallz}
\newcommand{\mzsq}{\mu_\smallz^2}
\newcommand{\mzc}{\mu_{\smallz 0}}
\newcommand{\cmz}{\bar{\mu}_{\smallz}}
\newcommand{\oa}{${\cal O}(\alpha)~$}
\newcommand{\oaa}{${\cal O}(\alpha^2)~$}
\newcommand{\oas}{${\cal O}(\alpha_s)~$}
\newcommand{\oaas}{${\cal O}(\alpha\alpha_s)~$}
\newcommand{\sineffl}{\sin\theta_{eff}^{\ell}\,}
\newcommand{\coseffl}{\cos\theta_{eff}^{\ell}\,}
\newcommand{\seffl}{\sin^2\theta_{eff}^{\ell}\,}
\newcommand{\ceffl}{\cos^2\theta_{eff}^{\ell}\,}
\newcommand{\sw}{s_\smallw\,}
\newcommand{\cw}{c_\smallw\,}
\newcommand{\swd}{s_\smallw^2\,}
\newcommand{\cwd}{c_\smallw^2\,}

\newcommand{\dis}{}
\newcommand{\overbar}[1]{mkern-1.5mu\overline{\mkern-1.5mu#1\mkern-1.5mu}\mkern
1.5mu}

\tikzset{
particle/.style={thick,draw=blue, postaction={decorate},
    decoration={markings,mark=at position .5 with {\arrow[blue]{triangle 45}}}},
uup/.style={draw=blue, postaction={decorate},
    decoration={markings,mark=at position .5 with {\arrow[blue]{stealth}}}},
top/.style={draw=red, postaction={decorate},
    decoration={markings,mark=at position .5 with {\arrow[red]{stealth}}}},
photon/.style={decorate, draw=black,
    decoration={coil,aspect=0,segment length=5pt,amplitude=1.5pt}},
Zboson/.style={decorate, draw=blue,
    decoration={coil,aspect=0,segment length=5pt,amplitude=1.5pt}},
Wboson/.style={decorate, draw=red,
    decoration={coil,aspect=0,segment length=5pt,amplitude=1.5pt}},
gluon/.style={decorate, draw=black,
    decoration={coil,aspect=0.5,segment length=2pt,amplitude=2pt}},
fermion/.style={draw=blue,
      postaction={decorate},decoration={markings,mark=at position .55
        with {\arrow[draw=blue]{>}}}},
hadron/.style={draw=black,
      postaction={decorate},decoration={markings,mark=at position .55
        with {\arrow[draw=black]{>}}}},
lepton/.style={draw=red,
      postaction={decorate},decoration={markings,mark=at position .55
        with {\arrow[draw=red]{>}}}},
neutrino/.style={draw=mygreen,
      postaction={decorate},decoration={markings,mark=at position .55
        with {\arrow[draw=mygreen]{>}}}},
vector/.style={decorate, decoration={snake}, draw},
crayon2b/.style={draw=blue!40!white, line width=2pt, line join=round,
  decoration={random steps, segment length=0.25pt, amplitude=0.5pt}, decorate},
crayon2g/.style={draw=green!40!white, line width=2pt, line join=round,
  decoration={random steps, segment length=0.25pt, amplitude=0.5pt}, decorate},
crayon2r/.style={draw=red!40!white, line width=2pt, line join=round,
  decoration={random steps, segment length=0.25pt, amplitude=0.5pt}, decorate},
crayon5/.style={line width=5pt, line join=round,
  decoration={random steps, segment length=0.5pt, amplitude=1.5pt}, decorate}
 }
%*******
% Intro
%*******

\section{Introduction}
\setcounter{equation}{0}
\label{sec:intro}

At hadron colliders, the production of a pair of leptons each with large transverse momentum, also known as Drell-Yan (DY) process \cite{Drell:1970wh}, has played a central role in the development of the Standard Model (SM) of the strong and electroweak (EW) interactions.
The same process that allowed the discovery of the $W$ and $Z$ bosons~\cite{Arnison:1983rp,Banner:1983jy,Arnison:1983mk,Bagnaia:1983zx}
provides today very precise information about the gauge sector of the SM and allows its test at the quantum level.
The determination of the masses of the gauge bosons and of the effective weak mixing angle, with relative errors at the 0.01\% and 0.1\% level respectively
\cite{Group:2012gb,Aaboud:2017svj,Aaltonen:2018dxj,ATLAS:2018gqq},
is possible thanks to the excellent quality of the data collected at the Fermilab Tevatron by the CDF and D0 experiments and at the CERN LHC by the ATLAS, CMS, and LHCb experiments.
The kinematical distributions are fitted by using theoretical templates, keeping the EW parameters as fitting variables.
The uncertainty in the evaluation of the templates enters  in the total error of the experimental value as a theoretical systematic contribution \cite{CarloniCalame:2016ouw,Bagnaschi:2019mzi,Behring:2021adr}.
Hence, it is of utmost importance to minimise the latter in order to make the comparison more significant.
The direct search for signals of New Physics is carried on, in the DY process, with the precise measurement of the tail of the kinematical distributions.
A deviation could then be interpreted as the distortion due to a new resonance or, in general, as the effect of a new higher-dimensional operator whose increment
with the energy shows up in the tails. In order to appreciate the possibility of such a deviation, the most precise SM prediction is needed, with the smallest possible residual theoretical uncertainty.
The inclusion of higher-order corrections makes the theoretical predictions more precise, but it requires to overcome challenging mathematical and computational problems.

The radiative corrections to the DY processes in the strong coupling $\as$
were computed
at next-to-leading-order~(NLO)~\cite{Altarelli:1979ub} and next-to-next-to-leading-order~(NNLO)~\cite{Hamberg:1990np,Harlander:2002wh} in perturbative Quantum Chromodynamics~(QCD).
Recently, the calculations at next-to-next-to-next-to-leading-order (N$^3$LO) QCD of the inclusive production of a virtual photon~\cite{Duhr:2020seh}, a $W$ boson~\cite{Duhr:2020sdp}, and for the total neutral current (NC) DY process \cite{Duhr:2021vwj} have been completed.
Fully differential NNLO QCD computations including the leptonic decay of the vector boson have been
presented in refs.~\cite{Anastasiou:2003yy,Anastasiou:2003ds,Melnikov:2006kv,Catani:2009sm,Catani:2010en},
and the first estimates of the fiducial cross sections for the NC DY process at  N$^3$LO QCD  have appeared~\cite{Camarda:2021ict}.
A plethora of papers discusses the inclusion of higher-order corrections in
the threshold region \cite{Moch:2005ky,Laenen:2005uz,Ravindran:2005vv,Ravindran:2006cg,deFlorian:2012za,Ahmed:2014cla,Catani:2014uta,Li:2014afw,Ajjath:2020ulr}.
The complete NLO EW corrections in the EW coupling $\alpha$ have been computed for $W$ production  in refs.~\cite{Dittmaier:2001ay,Baur:2004ig,Zykunov:2006yb,Arbuzov:2005dd,CarloniCalame:2006zq},
while the $Z$ production case has been considered in refs.~\cite{Baur:2001ze,Zykunov:2005tc,CarloniCalame:2007cd,Arbuzov:2007db,Dittmaier:2009cr}.
The NLO  QCD and NLO EW corrections are separately large, often reaching the ${\cal O}(10-20\%)$ of the leading order (LO) when we consider differential kinematical distributions,
raising the suspect that also mixed NNLO QCD-EW effects might be potentially large.
The fact that in specific phase-space corners NLO QCD and NLO EW corrections have opposite signs leads to large cancellations  \cite{Balossini:2009sa}.
When such cancellations take place, a reliable estimate of the residual uncertainties based on the previous orders is problematic, since in these regions we might expect non-negligible effects from the next perturbative orders.
Only an explicit calculation can solve the problem.

The combination of NLO QCD and EW results, consistently matched with QCD and QED Parton Showers,
has been presented in~\cite{Bernaciak:2012hj,Barze:2012tt,Barze:2013fru,Frederix:2018nkq}.
These Monte Carlo tools provide an approximation of the mixed QCD-EW corrections, valid in the soft and/or collinear limits for the radiation of multiple additional partons,
but can not assess the size of the genuine NNLO QCD-EW terms. Instead, they provide a rough estimate of the residual uncertainties.

Since high precision is required by the studies aiming at the determination of the EW parameters and the searches for New Physics,
an important effort of the theory community has been recently devoted to the evaluation of the mixed QCD-EW corrections.
The NNLO mixed QCD-QED corrections were obtained in ref.~\cite{deFlorian:2018wcj}
for the inclusive production of an on-shell $Z$ boson and extended at differential level for the decay of the $Z$ boson into a pair of neutrinos ~\cite{Cieri:2020ikq}.
The production of an on-shell $Z$ boson,
including NNLO mixed QCD-QED initial state corrections, but also
the factorised NLO QCD corrections to $Z$ production and the NLO QED corrections to the leptonic $Z$ decay,
has been discussed in ref.~\cite{Delto:2019ewv}.
The complete NNLO QCD-EW corrections at \oaas
in analytical form for the total cross section production of an on-shell $Z$ bosons have been presented
in refs.~\cite{Bonciani:2016wya,Bonciani:2019nuy,Bonciani:2020tvf,Bonciani:2021iis},
whereas results differential in the final-state leptonic variables
have been presented in refs.~\cite{Buccioni:2020cfi} and ~\cite{Behring:2020cqi}
for the on-shell $Z$ and $W$ production respectively.
Beyond the on-shell approximation, results have been obtained in the {\it pole} approximation \cite{Denner:2019vbn},
which offers the possibility to move beyond the strictly on-shell limit
and is based on a systematic expansion around the $W$ or $Z$ resonance. By definition, its range of validity is expected to be restricted to this phase-space region.
The dominant part of the mixed QCD-EW corrections at the $W$ or $Z$ resonances has been evaluated with this technique in refs.~\cite{Dittmaier:2014qza,Dittmaier:2015rxo}.
Precise predictions at NNLO QCD-EW level in the whole phase-space have been obtained in ref.~\cite{Dittmaier:2020vra} for the subset of ${\cal O}(n_F\as \alpha)$.
The charged-current DY process $pp\to \ell \nu_\ell+X$, has been considered in ref.~\cite{Buonocore:2021rxx} including the mixed QCD-EW corrections,
with the reweighted two-loop virtual corrections in the {\it pole} approximation
and all the other contributions in exact form.
Very recently the complete set of NNLO QCD-EW corrections to the NC DY process $pp\to \ell^+\ell^-+X$
has been presented in ref.~\cite{Bonciani:2021zzf}, with the inclusion of the exact two-loop virtual
contributions\footnote{The impact of the IR subtraction technique adopted in ref.~\cite{Bonciani:2021zzf}
has been presented in refs.~\cite{Buonocore:2021tke, Camarda:2021jsw}.}.

The virtual contributions were one of the bottlenecks for a complete ${\cal O}(\as\alpha)$ calculation, because
the evaluation of the $2\to 2$ two-loop Feynman diagrams with internal masses is at the frontier of current computational techniques.
Progress on the evaluation of the corresponding two-loop master integrals has been reported in refs.~\cite{Bonciani:2016ypc,Heller:2019gkq,Hasan:2020vwn,Long:2021fdc}
and, recently,
the computation of the two-loop helicity amplitudes for NC massless lepton pair production was discussed in ref.~\cite{Heller:2020owb}.
In this article, we provide in detail the independent computation at \oaas of the two-loop amplitude for the process $q\bar q\to\ell^+\ell^-$.
The calculation considers a massive leptonic final state,
and features collinear logarithms of the lepton mass.
These results, combined with the remaining perturbative contributions,
allow the numerical evaluation of the hadron-level cross-section at this perturbative order,
as presented in \cite{Bonciani:2021zzf}.

In Section \ref{sec:frame},
we provide the framework of our computation, with the description of
the treatment of $\gamma_5$ and of the ultraviolet (UV)
renormalisation procedure.
In Section \ref{sec:ampl},
the details of the computation are presented,
from the amplitude level until the renormalised matrix elements.
In Section \ref{sec:infra},
the infrared (IR) subtraction procedure is discussed.
In Section \ref{sec:results}
we describe how we obtain the UV renormalised IR subtracted finite remainder
and we present the details necessary for the numerical evaluation
of the analytical expressions.
In Section \ref{sec:conclusions}
we draw our conclusions.
The unrenormalised matrix elements in terms of master integrals,
together with a benchmark grid for the finite reminder,
are provided as ancillary material.
Appendix \ref{app:readme} gives the reader the information needed to make use of the files.

% ******* Theory *******
%%%%%%%%%%%%%%%%%%%%%%%%%%%%%%%%%%%%%%%%%%%%%%%%%%%%%%%%%%%%%%%%%%%%%%%%%%%%
\section{Framework of the Calculation}
\label{sec:frame}
\subsection{The process}
We consider two-loop mixed QCD-EW corrections to lepton pair production in the quark annihilation channel, as given by
\begin{equation}
 q(p_1) + \bar{q}(p_2) \rightarrow l^{-}(p_3) + l^{+} (p_4) \,.
\label{eq:process}
\end{equation}
      We have separately computed the results for the $u\bar u$ and $d\bar d$ channels,
      with $u$ and $d$ denoting generic massless up- and down-type quarks.
      In the paper we present explicit results for the $u\bar u$ channel.
The bare amplitude\footnote{The \textit{hat} denotes the bare quantities.}
for this partonic process admits a double perturbative expansion in the two coupling constants
\begin{align}
 |\unM \rangle = |\unM^{(0)} \rangle + \asrhat |\unM^{(1,0)} \rangle + \hat\aem |\unM^{(0,1)} \rangle + \asrhat \hat\aem |\unM^{(1,1)} \rangle + \cdots
\end{align}
In this paper, we present the calculation of the following interference terms:
\begin{equation}
  \langle \unM^{(0)} | \unM^{(1,0)} \rangle \,, ~~
  \langle \unM^{(0)} | \unM^{(0,1)} \rangle \,, ~~
  \langle \unM^{(0)} | \unM^{(1,1)} \rangle \,,
  \label{eq:interferences}
\end{equation}
which contribute at \oaas to the unpolarised squared matrix element
of the process in eq.~\eqref{eq:process}.
These scalar quantities contain many one- and two-loop Feynman integrals.
We apply the
integration-by-parts (IBP) \cite{Tkachov:1981wb,Chetyrkin:1981qh} and
Lorentz invariance (LI) \cite{Gehrmann:1999as} identities
to reduce those scalar Feynman integrals to a limited set, the so-called master integrals (MIs).
The corrections in eq.~(\ref{eq:interferences}) can then be expressed as the sum of the MIs, each
multiplied by its corresponding coefficient.
The latter are rational functions of the kinematical invariants
and of the masses which appear in the process.

Virtual corrections are affected by singularities both of UV and IR type.
The renormalisation procedure allows us to cancel the UV divergences
and to express the amplitude in terms of renormalised parameters,
which are in turn related to measurable quantities.
The presence of IR divergences in the virtual corrections, on the other hand, is the counterpart of those
which appear upon phase-space integration in the processes with the emission of additional massless partons,
in our case a photon and/or a gluon.
The universality of the IR divergent structure of the scattering amplitudes arising from soft and/or collinear radiation
has been widely discussed in the literature and
allows us to build a subtraction term, independently of the details of the full two-loop calculation.
We exploit this property to check the consistency of our computation,
in particular for the prescription related to the Dirac $\gamma_5$ matrix.
To regularise both UV and IR divergences, we work in $d=4-2\varepsilon$ space-time dimensions.
The renormalised amplitude can thus be written as a Laurent expansion in $\varepsilon$.
At one-loop level we have:
\bea
  \langle {\cal M}^{(0)} | {\cal M}^{(0,1)} \rangle \, &=& \sum_{i=-2}^{2} \varepsilon^i C^{(0,1)}_i(s,t,\mw,\mz,\ml)\, ,\label{i01}\\
  \langle {\cal M}^{(0)} | {\cal M}^{(1,0)} \rangle \, &=& \sum_{i=-2}^{2} \varepsilon^i C^{(1,0)}_i(s,t,\mw,\mz,\ml)\, ,\label{i10}
\eea
while at two-loop we have:
\bea
  \langle {\cal M}^{(0)} | {\cal M}^{(1,1)} \rangle \, &=& \sum_{i=-4}^{0} \varepsilon^i C^{(1,1)}_i(s,t,\mw,\mz,\ml)\,\label{i11} .
\eea
      In eq.~\ref{i11} we have retained only terms up to $\varepsilon^0$,
      as we are only interested in the finite cotributions at NNLO.
      However, this series contains higher powers of $\varepsilon$
      which will be relevant for higher-order calculations.
In these expressions
$m_l$ is the lepton mass, and $\mw$, $\mz$ are the complex masses
of the $W$ and $Z$ boson
defined as:
\be
\mu_V^2=M_V^2-i M_V\Gamma_V\;.
\ee
The mass and decay width $M_V,\Gamma_V$
are real parameters and are the pole quantities,
i.e. correspond to the position of the pole in the complex plane of the gauge boson propagator.
We also introduce the Mandelstam variables, defined as follows:
\begin{equation}
 s = (p_1+p_2)^2, \, t = (p_1-p_3)^2, \, u = (p_2-p_3)^2 \,\, {\rm with} \,\, s+t+u=2 m_l^2 \,.
\end{equation}
The on-shell conditions of the external particles are given by
\begin{equation}
 p_1^2 = p_2^2 = 0; ~ p_3^2 = p_4^2 = m_l^2;
\end{equation}
The coefficients $C_i$ in eq.~(\ref{i11}) constitute the IR-unsubtracted two-loop interference terms,
while those in eqs.~(\ref{i01},\ref{i10}), computed at one-loop,
are necessary to prepare the subtraction term.
The IR-subtracted expressions that can be obtained by their combination constitutes the main result that we present in this paper.

\subsection{Treatment of \texorpdfstring{$\gamma_5$}{g5} in Dimensional Regularisation}
\label{sec:gamma5}

The calculation of radiative corrections, including chiral quantities in dimensional regularisation, faces the problem of defining a generalisation of
the inherently four-dimensional object  $\gamma_5$ in the arbitrary space-time dimension $d$.
In \cite{'tHooft:1972fi}, 't Hooft and Veltman proposed to abandon, in $d$ dimensions, the anticommutation relation of $\gamma_5$
\begin{equation}
 \{\gamma_\mu,\gamma_5\} = 0
\end{equation}
and to keep the cyclicity of the Dirac traces; the price is a significant
computational load, due to the different treatment of the four-momenta
components defined in the four space-time dimensions and of those
in the remaining $d-4$.
In refs.~\cite{Kreimer:1989ke,Korner:1991sx}
Kreimer et al. retained the anticommutation relation, with a substantial computational gain, and abandoned the cyclicity property of the trace operation.
It was demonstrated that, at least in one-loop calculations, the choice of
a unique point to start the evaluation of  all the traces needed in one calculation
leads to the systematic cancellation of the prescription dependent terms,
along with that of the IR poles.
It has recently been explicitly checked in ref.~\cite{Heller:2020owb},
for the NC DY process at \oaas\!,
that the implementations at two-loop level of the 't Hooft-Veltman-Breitenloner-Mason and of the Kreimer prescriptions
yield, as expected, different results for the scattering amplitudes,
but are in perfect agreement when one considers the finite correction
after subtraction of all the IR divergences.

In this computation, we retain the anticommutation relation together with
\begin{equation}
 \gamma_5^2 = 1 \,,  ~~~ \gamma_5^{\dag} = \gamma_5
\end{equation}
and keep a fixed point to write and then evaluate the Dirac traces.
As a result, we end up with zero or one $\gamma_5$ at the rightmost position of the Dirac trace.
In the single $\gamma_5$ case, we perform the replacement
\begin{align}
  \gamma_5 = i \frac{1}{4!} \varepsilon_{\nu_1 \nu_2 \nu_3 \nu_4}
  \gamma^{\nu_1}  \gamma^{\nu_2} \gamma^{\nu_3} \gamma^{\nu_4} \,,
\end{align}
where, $\varepsilon^{\mu\nu\rho\sigma}$ is the Levi-Civita tensor.
The contraction of two such tensors yields
\begin{align}
  \label{eqn:LeviContract}
  \varepsilon_{\mu_1\nu_1\lambda_1\sigma_1}\,\varepsilon^{\mu_2\nu_2\lambda_2\sigma_2}=\,-\,
  {\left |
  \begin{array}{cccc}
    \delta_{\mu_1}^{\mu_2} &\delta_{\mu_1}^{\nu_2}&\delta_{\mu_1}^{\lambda_2} & \delta_{\mu_1}^{\sigma_2}\\
    \delta_{\nu_1}^{\mu_2}&\delta_{\nu_1}^{\nu_2}&\delta_{\nu_1}^{\lambda_2}&\delta_{\nu_1}^{\sigma_2}\\
    \delta_{\lambda_1}^{\mu_2}&\delta_{\lambda_1}^{\nu_2}&\delta_{\lambda_1}^{\lambda_2}&\delta_{\lambda_1}^{\sigma_2}\\
    \delta_{\sigma_1}^{\mu_2}&\delta_{\sigma_1}^{\nu_2}&\delta_{\sigma_1}^{\lambda_2}&\delta_{\sigma_1}^{\sigma_2}
  \end{array}
                                                                                       \right |}
\end{align}
and all the Lorentz indices are considered to be $d$-dimensional.

The presence of a prescription-dependent term of ${\cal O}(\varepsilon)$ in the squared matrix element
affects all the coefficients in the Laurent expansion, with the exception of the highest pole:
in fact,
the product of a term of ${\cal O}(\varepsilon)$ with a singular factor $\varepsilon^{-k}$, with $k>0$,
generates a contribution of ${\cal O}(\varepsilon^{-k+1})$.
Such prescription-dependent terms will be generated both in the unsubtracted squared matrix element and in the subtraction term.
The cancellation of the IR singularities,
expected on general grounds,
requires that also the prescription-dependent terms cancel accordingly.
In the present calculation, the IR subtraction term is computed by following the properties of universality of the radiation in the IR limits,
combining the universal divergent structure with the Born and one-loop amplitudes.
The construction of this subtraction term is completely independent with respect to the evaluation of the two-loop amplitude and it
provides a non-trivial check of our algebraic manipulations.
We observe the cancellation of all the lower order poles, when combining the full two-loop amplitude with the subtraction term, which hints in favour of the consistency of our approach.

\subsection{Ultraviolet renormalisation}
\label{sec:UV}
The renormalisation at \oaas of the NC DY process has already been discussed in detail in ref.~\cite{Dittmaier:2020vra}.
We report here the basic steps that we have implemented to obtain the complete two-loop renormalised amplitude.
\subsubsection{Charge renormalisation}
The bare gauge couplings $g_0,g'_0$ and the Higgs doublet vacuum expectation value $v_0$ are
expressed in terms of their renormalised counterparts $g,g',v$
via the introduction of appropriate counterterms.
The relation of $g,g',v$ to a set of three measurable quantities,
like for instance $G_\mu,\mw,\mz$ (with $G_\mu$ the Fermi constant)
or $\alpha,\mw,\mz$ (with $\alpha$ the fine structure constant)\footnote{
  An alternative input scheme, relevant for the direct determination of the weak mixing angle,
  has been presented in ref.~\cite{Chiesa:2019nqb}
  },
eventually allows the numerical evaluation of the amplitude.
We define for convenience two additional bare quantities:
the sine squared of the on-shell weak mixing angle,
which we abbreviate as $s_{\smallw 0}^2=\sin^2\theta_{\smallw0}=1-\frac{\mu_{\smallw0}^2}{\mu_{\smallz 0}^2}$, $c_{\smallw 0}^2=1-s_{\smallw 0}^2$, and
the electric charge $e_0=g_0 s_{\smallw 0}$, but we stress that only three of the above parameters are independent.
We introduce the relation between bare and renormalised input parameters
\bea
\mwc^2&=&\mwsq+\delta\mwsq,\quad  \mzc^2=\mzsq+\delta\mzsq,\quad e_0=e+\delta e\, ,
\label{eq:dmass}
 \\
\frac{\delta \swd}{\swd}&=&
\frac{\cwd}{\swd}\left( \frac{\delta\mzsq}{\mzsq}-\frac{\delta\mwsq}{\mwsq} \right)\, .
\label{eq:ds2s2}
\eea
The on-shell electric charge counterterm $\delta e$ at \oaa and \oaas has been discussed in ref.~\cite{Degrassi:2003rw},
from the study of the Thomson scattering.
The mass counterterms $\delta\mwsq,\, \delta\mzsq$ in the complex mass scheme  \cite{Denner:2005fg}
have been presented in ref.~\cite{Dittmaier:2020vra}.
In terms of the transverse part of the unrenormalised $VV$ gauge boson self-energies, they are defined as follows:
\be
\delta\mu_V^2 = \Sigma_{T}^{VV}(\mu_V^2)\, ,
\ee
at the pole in the complex plane $q^2=\mu_V^2$ of the gauge boson propagator, with $V=W,Z$.
From the study of the muon-decay amplitude, we derive the following relation
\be
\frac{G_\mu}{\sqrt{2}}=
\frac{\pi \alpha}{2 \mwsq \swd}\left( 1+ \Delta r \right)\, .
\ee
The finite correction $\Delta r$ was introduced,
with real gauge boson masses, in ref.~\cite{Sirlin:1980nh}
and its \oaas corrections were presented in ref.~\cite{Kniehl:1989yc,Djouadi:1993ss}. We evaluate it here with complex-valued masses.

We consider now the bare couplings which appear at tree-level in the interaction of the photon and $Z$ boson with fermions.
The UV divergent correction factors $\delta g_Z^\alpha$ contributes to the
charge renormalisation of the $Zf\bar f$ vertex in the  $(\alpha,\mw,\mz)$ input scheme
\bea
\frac{g_0}{ c_{\smallw 0}} &=& \frac{e}{\cw \sw}
\left[ 1
+\frac12 \left( 2 \frac{\delta e}{e}
+\frac{\swd-\cwd}{\cwd}\frac{\delta \swd}{\swd}
\right)
\right]
\equiv
\frac{\sqrt{4\pi\alpha}}{\cw \sw} \left(
1 +
\delta g_Z^\alpha \right)\, ,
\label{eq:shiftsZalpha}
\eea
while  $\delta g_Z^{G_\mu}$ is relevant in the  $(G_\mu,\mw,\mz)$ input scheme
\bea
\frac{g_0}{ c_{\smallw 0}} &=&
\sqrt{4\sqrt{2} G_\mu \mz^2}
\left[ 1- \frac12 \Delta r
+\frac12 \left( 2 \frac{\delta e}{e}
+\frac{\swd-\cwd}{\cwd}\frac{\delta \swd}{\swd}
\right)
\right]
\equiv
\sqrt{4\sqrt{2} G_\mu \mz^2} \left(
1 + \delta g_Z^{G_\mu} \right)~~~~~~~~~
\label{eq:shiftsZGmu}
\eea
Working out the explicit expression of $\delta g_Z^{G_\mu}$, the dependence on the electric charge counterterm cancels out.
Analogously, in the case of the $\gamma f \bar f$ vertex,
the electric charge renormalisation is given by
\bea
g_0\, s_{\smallw 0} &= e_0
= e+\delta e
\equiv \sqrt{4\pi\alpha} \left(1 + \delta g_A^\alpha \right)
\label{eq:shiftsgammaalpha}
\eea
in the  $(\alpha,\mw,\mz)$  scheme and by
\bea
g_0\, s_{\smallw 0} &=
\sqrt{4\sqrt{2} G_\mu \mw^2 \swd}
\left[  1+\frac12 \left(-\Delta r+2\frac{\delta e}{e} \right)  \right]
\equiv
\sqrt{4\sqrt{2} G_\mu \mw^2 \swd}
\left( 1 + \delta g_A^{G_\mu} \right)
\label{eq:shiftsgammaGmu}
\eea
in the $(G_\mu,\mw,\mz)$ scheme.
The counterterm contributions to the renormalised amplitude are obtained by replacing
the bare couplings in the lower order amplitudes with the expressions
presented in eqs.~(\ref{eq:shiftsZalpha}-\ref{eq:shiftsgammaGmu})
and expanding $\delta g_{A,Z}$ up to the relevant perturbative order.
We show in the next Section how  $\delta g_{A,Z}$ enter in the renormalisation of the gauge boson propagators.

\subsubsection{Renormalisation of the gauge boson propagators}
\label{sec:gaugeren}
The renormalised gauge boson self-energies are obtained, at \oa\hspace{-0.5em},
by combining the unrenormalised self-energy expressions with the mass and wave function counterterms.
In the full calculation, we never introduce wave function counterterms on the internal lines,
because they would systematically cancel
against a corresponding factor stemming from the definition of the renormalised vertices.
We exploit instead the relation in the SM
between the wave function and charge counterterms \cite{Denner:2019vbn}
and we directly use the latter to define the renormalised self-energies.
We obtain:
\bea
\Sigma_{R,T}^{AA}(q^2) &=&
\Sigma_T^{AA}(q^2) + 2 \,q^2\,\delta g_A\, , \\
\Sigma_{R,T}^{ZZ}(q^2) &=&
\Sigma_T^{ZZ}(q^2) -\delta\mu_Z^2 + 2\, (q^2-\mzsq)\,\delta g_Z \, ,\\
\Sigma_{R,T}^{AZ}(q^2) &=&
\Sigma_T^{AZ}(q^2)  - q^2\,\frac{\delta \swd}{\sw\cw} \, , \\
\Sigma_{R,T}^{ZA}(q^2) &=&
\Sigma_T^{ZA}(q^2)  - q^2\,\frac{\delta \swd}{\sw\cw},
\eea
where $\Sigma_T^{VV}$ and $\Sigma_{R,T}^{VV}$ are the transverse part of the bare and renormalised $VV$ vector boson self-energy.
The charge counterterms have been defined in eqs.~(\ref{eq:shiftsZalpha}-\ref{eq:shiftsgammaGmu}).
At \oaas the structure of these contributions does not change:
the corrections to the gauge boson self-energies
stem from a quark loop with one internal gluon exchange and,
in addition, from the \oas mass renormalisation of the quark lines in the one-loop self-energies.

The expression of the two-loop Feynman integrals required for the evaluation of the
\oaas correction to the gauge boson propagators and all the needed counterterms
can be found in refs.~\cite{Kniehl:1989yc,Djouadi:1993ss,Dittmaier:2020vra}.

\subsubsection{Checks}
The physical scattering amplitude does not depend on the gauge choice used to quantise the theory.
In our computation we use the background field gauge (BFG) \cite{Denner:1994xt}, which restores some QED-like Ward identities in the full EW SM
and guarantees in turn some properties of the Green's function which appear in the calculation.
For instance, the UV finiteness of the vertex corrections, summed with the external wave function of the fermions entering in that vertex, is an important property which can provide an additional check of the calculation.
This check has been performed explicitly at the one-loop level, based on the knowledge of the UV poles of the one-loop Feynman integrals.
At two-loop level, we consider the UV finiteness of the box corrections,
we explicitly renormalise the tree-level propagators, and we assume
that the cancellation of the UV poles between vertex and external wave function corrections takes place.
Since the propagator corrections are IR finite,
we conclude that the sum of vertex, box, and external wave function corrections contains only IR poles.
We have explicitly checked that the coefficients of these poles
exactly match the ones predicted by the universal structure
of the IR divergences in gauge theories \cite{Catani:1998bh}.
This result hints in favour of the validity of the Ward identity
also at two-loop level.

%%%%%%%%%%%%%%%%%%%%%%%%%%%%%%%%%%%%%%%%%%%%%%%%%%%%%%%%%%%%%%%%%%%%%%%%%%%%%
\section{The UV-renormalised unsubtracted amplitude}
\label{sec:ampl}
\subsection{Generation of the full amplitude and evaluation of the interference terms}
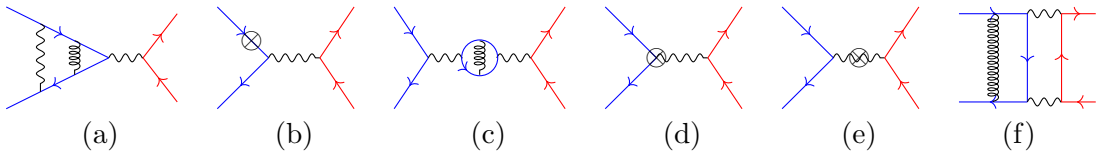
\begin{figure}[ht]
\centering
\begin{tikzpicture}[scale=0.45]
 \draw[fermion] (-3.0,1.5) -- (0,0);
 \draw[fermion] (0,0) -- (-3.0,-1.5);
 \draw[photon] (-2.0,1.0) -- (-2.0,-1.0);
 \draw[gluon] (-1.0,0.5) -- (-1.0,-0.5);
 \draw[photon] (0,0) -- (1,0);
 \draw[lepton] (2,-1.3) -- (1.0,0);
 \draw[lepton] (1.0,0) -- (2,1.3);
\node at (-0.25,-2.3) {(a)};
\end{tikzpicture}
\quad
\begin{tikzpicture}[scale=0.45]
 \draw[fermion] (-3.0,1.5) -- (-1.5,0);
 \draw[fermion] (-1.5,0) -- (-3.0,-1.5);
 \draw[photon] (-1.5,0) -- (0,0);
%  \draw[photon] (0,0) -- (1,0);
 \draw[lepton] (1,-1.5) -- (0,0);
 \draw[lepton] (0,0) -- (1,1.5);
 \node at (-2.0,0.5) {$\otimes$};
 \node at (-0.75,-2.3) {(b)};
\end{tikzpicture}
\quad
\begin{tikzpicture}[scale=0.45]
 \draw[fermion] (-3.0,1.5) -- (-2.0,0);
 \draw[fermion] (-2.0,0) -- (-3.0,-1.5);
 \draw[gluon] (-0.5,0.5) -- (-0.5,-0.5);
 \draw[photon] (-2,0) -- (-1,0);
 \draw[photon] (0,0) -- (1,0);
 \draw[lepton] (2,-1.5) -- (1.0,0);
 \draw[lepton] (1.0,0) -- (2,1.5);
\draw[blue, decoration={markings, mark=at position 0.625 with {\arrow{>}}},
        postaction={decorate}] (-0.5,0) circle (15pt);
\node at (-0.25,-2.3) {(c)};
\end{tikzpicture}
\quad
\begin{tikzpicture}[scale=0.45]
 \draw[fermion] (-3.0,1.5) -- (-1.5,0);
 \draw[fermion] (-1.5,0) -- (-3.0,-1.5);
 \draw[photon] (-1.5,0) -- (0,0);
 \draw[lepton] (1,-1.5) -- (0,0);
 \draw[lepton] (0,0) -- (1,1.5);
 \node at (-1.5,0) {$\otimes$};
 \node at (-0.75,-2.3) {(d)};
\end{tikzpicture}
\quad
\begin{tikzpicture}[scale=0.45]
 \draw[fermion] (-3.0,1.5) -- (-1.5,0);
 \draw[fermion] (-1.5,0) -- (-3.0,-1.5);
 \draw[photon] (-1.5,0) -- (0,0);
 \draw[lepton] (1,-1.5) -- (0,0);
 \draw[lepton] (0,0) -- (1,1.5);
 \node at (-0.75,0) {$\otimes$};
 \node at (-0.75,-2.3) {(e)};
\end{tikzpicture}
\quad
\begin{tikzpicture}[scale=0.45]
 \draw[fermion] (-2.0,1.3) -- (0,1.3);
 \draw[fermion] (0,1.3) -- (0,-1.3);
 \draw[fermion] (0,-1.3) -- (-2.0,-1.3);
 \draw[gluon] (-1.0,1.3) -- (-1.0,-1.3);
 \draw[photon] (0.0,1.3) -- (1,1.3);
 \draw[photon] (0,-1.3) -- (1,-1.3);
 \draw[lepton] (1,1.3) -- (2,1.3);
 \draw[lepton] (1,-1.3) -- (1,1.3);
 \draw[lepton] (2,-1.3) -- (1,-1.3);
\node at (-0.25,-2.3) {(f)};
\end{tikzpicture}
\caption{Sample Feynman diagrams of two-loop corrections and associated two-loop counterterms.
}
\label{fig:diagsamples}
\end{figure}

\begin{figure}[ht]
\centering
\begin{tikzpicture}[scale=0.45]
 \draw[fermion] (-3.0,1.5) -- (-1.0,0);
 \draw[fermion] (-1.0,0) -- (-3.0,-1.5);
 \draw[gluon] (-2.0,0.7) -- (-2.0,-0.7);
 \draw[photon] (1.0,0.7) -- (1.0,-0.7);
 \draw[photon] (0,0) -- (-1,0);
 \draw[lepton] (2,-1.5) -- (0.0,0);
 \draw[lepton] (0.0,0) -- (2,1.5);
\node at (-0.25,-2.3) {(a)};
\end{tikzpicture}
\quad
\begin{tikzpicture}[scale=0.45]
 \draw[fermion] (-3.0,1.5) -- (-1.0,0);
 \draw[fermion] (-1.0,0) -- (-3.0,-1.5);
 \draw[gluon] (-1.8,0.6) -- (-1.8,-0.6);
 \draw[photon] (-1,0) -- (0,0);
 \draw[lepton] (1.5,-1.5) -- (0,0);
 \draw[lepton] (0,0) -- (1.5,1.5);
 \node at (1,1.1) {$\otimes$};
 \node at (-0.5,-2.3) {(b)};
\end{tikzpicture}
\quad
\begin{tikzpicture}[scale=0.45]
 \draw[fermion] (-3.0,1.5) -- (-1.2,0);
 \draw[fermion] (-1.2,0) -- (-3.0,-1.5);
 \draw[gluon] (-2.0,0.6) -- (-2.0,-0.6);
 \draw[photon] (-0.5,0) -- (-1.2,0);
 \draw[photon] (0.2,0) -- (1,0);
 \draw[lepton] (2,-1.5) -- (1.0,0);
 \draw[lepton] (1.0,0) -- (2,1.5);
\draw[blue, decoration={markings, mark=at position 0.625 with {\arrow{>}}},
        postaction={decorate}] (-0.16,0) circle (10pt);
\node at (-0.25,-2.3) {(c)};
\end{tikzpicture}
\quad
\begin{tikzpicture}[scale=0.45]
 \draw[fermion] (-3.0,1.5) -- (-1.0,0);
 \draw[fermion] (-1.0,0) -- (-3.0,-1.5);
 \draw[gluon] (-1.8,0.6) -- (-1.8,-0.6);
 \draw[photon] (-1,0) -- (0,0);
 \draw[lepton] (1,-1.5) -- (0,0);
 \draw[lepton] (0,0) -- (1,1.5);
 \node at (-1.0,0) {$\otimes$};
 \node at (-1,-2.3) {(d)};
\end{tikzpicture}
\quad
\begin{tikzpicture}[scale=0.45]
 \draw[fermion] (-3.0,1.5) -- (-1.0,0);
 \draw[fermion] (-1.0,0) -- (-3.0,-1.5);
 \draw[gluon] (-1.8,0.6) -- (-1.8,-0.6);
 \draw[photon] (-1,0) -- (0,0);
 \draw[lepton] (1,-1.5) -- (0,0);
 \draw[lepton] (0,0) -- (1,1.5);
 \node at (-0.5,0) {$\otimes$};
 \node at (-1,-2.3) {(e)};
\end{tikzpicture}
\quad
\begin{tikzpicture}[scale=0.45]
 \draw[fermion] (-3.0,1.5) -- (-1.0,0);
 \draw[fermion] (-1.0,0) -- (-3.0,-1.5);
 \draw[gluon] (-1.8,0.6) -- (-1.8,-0.6);
 \draw[photon] (-1,0) -- (0,0);
 \draw[lepton] (1,-1.5) -- (0,0);
 \draw[lepton] (0,0) -- (1,1.5);
 \node at (-2.4,1.1) {$\otimes$};
 \node at (-1,-2.3) {(f)};
\end{tikzpicture}
\caption{Sample Feynman diagrams of factorisable corrections, including one-loop counterterm corrections.}
\label{fig:diagsamplesCT}
\end{figure}
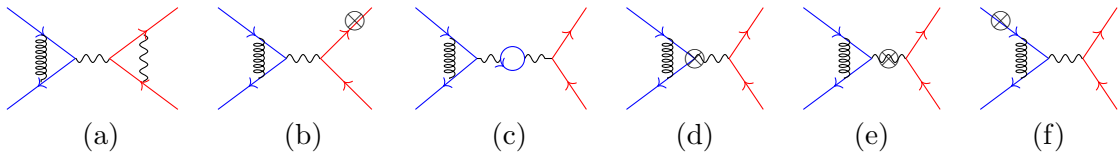

The amplitude at \oaas receives contributions from different kinds of Feynman diagrams, for which
%% %
we depict\footnote{
The straight line with an arrow represents a fermion line
(blue: massless quark, red: massive leptons, green: neutrinos),
the wavy lines represent EW gauge bosons
and the spiral coils represent gluons.
}
in Figures \ref{fig:diagsamples} and \ref{fig:diagsamplesCT}
a few representative examples.
In the initial state we have
two-loop vertex corrections (Figure~\ref{fig:diagsamples}-a),
which are combined with
the two-loop external quark wave function corrections
(Figure~\ref{fig:diagsamples}-b)
and with
the one-loop external quark wave function correction
with initial-state QCD vertex correction (Figure~\ref{fig:diagsamplesCT}-f),
yielding an UV-finite, but still IR-divergent, result.
The two-loop gauge boson self-energies,
together with the corresponding two-loop mass counterterms
and with the charge renormalisation constants of the initial and final state vertices
are shown in Figures (\ref{fig:diagsamples}-c,d,e)
and yield an UV- and IR-finite contribution.
An example of two-loop boxes with the exchange of neutral or charged EW bosons
is given in Figure \ref{fig:diagsamples}-f.
At \oaas all the factorizable contributions include an initial-state QCD vertex correction;
the second factor can be, alternatively:
the final-state EW vertex corrections,
the external lepton wave function correction,
the one-loop self-energy corrections,
the one-loop mass and charge renormalization counterterms, and
the one-loop external quark wave function correction.
They are schematically represented in
Figures~\ref{fig:diagsamplesCT}(a)--(f) respectively.
Also in this case the BFG properties allow to identify UV-finite combinations of Feynman diagrams.

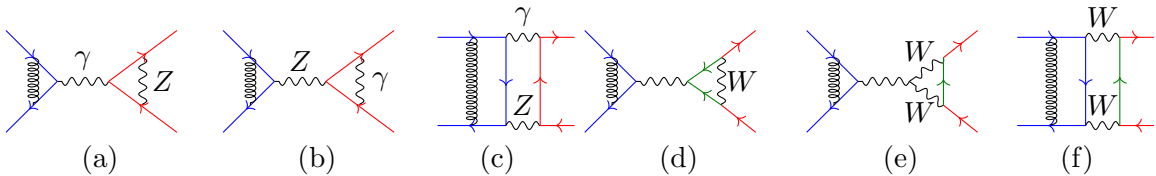
\begin{figure}[ht]
\centering
\begin{tikzpicture}[scale=0.45]
 \draw[fermion] (-3.0,1.5) -- (-1.5,0);
 \draw[fermion] (-1.5,0) -- (-3.0,-1.5);
 \draw[gluon] (-2.2,0.7) -- (-2.2,-0.7);
 \draw[photon] (1.0,0.7) -- (1.0,-0.7);
 \draw[photon] (0,0) -- (-1.5,0);
 \draw[lepton] (2,-1.5) -- (0.0,0);
 \draw[lepton] (0.0,0) -- (2,1.5);
\node at (-0.75,0.6) {$\gamma$};
\node at (1.6,0) {$Z$};
\node at (-0.25,-2.3) {(a)};
\end{tikzpicture}
\quad
\begin{tikzpicture}[scale=0.45]
 \draw[fermion] (-3.0,1.5) -- (-1.5,0);
 \draw[fermion] (-1.5,0) -- (-3.0,-1.5);
 \draw[gluon] (-2.2,0.7) -- (-2.2,-0.7);
 \draw[photon] (1.0,0.7) -- (1.0,-0.7);
 \draw[photon] (0,0) -- (-1.5,0);
 \draw[lepton] (2,-1.5) -- (0.0,0);
 \draw[lepton] (0.0,0) -- (2,1.5);
\node at (-0.75,0.6) {$Z$};
\node at (1.6,0) {$\gamma$};
\node at (-0.25,-2.3) {(b)};
\end{tikzpicture}
\quad
\begin{tikzpicture}[scale=0.45]
 \draw[fermion] (-2.0,1.3) -- (0,1.3);
 \draw[fermion] (0,1.3) -- (0,-1.3);
 \draw[fermion] (0,-1.3) -- (-2.0,-1.3);
 \draw[gluon] (-1.0,1.3) -- (-1.0,-1.3);
 \draw[photon] (0.0,1.3) -- (1,1.3);
 \draw[photon] (0,-1.3) -- (1,-1.3);
 \draw[lepton] (1,1.3) -- (2,1.3);
 \draw[lepton] (1,-1.3) -- (1,1.3);
 \draw[lepton] (2,-1.3) -- (1,-1.3);
\node at (0.5,-0.7) {$Z$};
\node at (0.5,1.9) {$\gamma$};
\node at (-0.25,-2.3) {(c)};
\end{tikzpicture}
\begin{tikzpicture}[scale=0.45]
 \draw[fermion] (-3.0,1.5) -- (-1.5,0);
 \draw[fermion] (-1.5,0) -- (-3.0,-1.5);
 \draw[gluon] (-2.2,0.7) -- (-2.2,-0.7);
 \draw[photon] (0,0) -- (-1.5,0);
 \draw[photon] (1.0,0.7) -- (1.0,-0.7);
 \draw[lepton] (2,-1.5) -- (1.0,-0.7);
 \draw[lepton] (1.0,0.7) -- (2,1.5);
 \draw[neutrino] (1.0,0.7) -- (0.0,0);
 \draw[neutrino] (1.0,-0.7) -- (0.0,0);
\node at (1.6,0) {$W$};
\node at (-0.25,-2.3) {(d)};
\end{tikzpicture}
\quad
\begin{tikzpicture}[scale=0.45]
 \draw[fermion] (-3.0,1.5) -- (-1.5,0);
 \draw[fermion] (-1.5,0) -- (-3.0,-1.5);
 \draw[gluon] (-2.2,0.7) -- (-2.2,-0.7);
\draw[photon] (0,0) -- (-1.5,0);
  \draw[photon] (1.0,0.7) -- (0,0);
  \draw[photon] (1.0,-0.7) -- (0,0);
 \draw[neutrino] (1.0,-0.7) -- (1.0,0.7);
  \draw[lepton] (2,-1.5) -- (1.0,-0.7);
 \draw[lepton] (1.0,0.7) -- (2,1.5);
\node at (0.3,0.9) {$W$};
\node at (0.3,-0.9) {$W$};
\node at (-0.25,-2.3) {(e)};
\end{tikzpicture}
\quad
\begin{tikzpicture}[scale=0.45]
 \draw[fermion] (-2.0,1.3) -- (0,1.3);
 \draw[fermion] (0,1.3) -- (0,-1.3);
 \draw[fermion] (0,-1.3) -- (-2.0,-1.3);
 \draw[gluon] (-1.0,1.3) -- (-1.0,-1.3);
 \draw[photon] (0.0,1.3) -- (1,1.3);
 \draw[photon] (0,-1.3) -- (1,-1.3);
 \draw[lepton] (1,1.3) -- (2,1.3);
 \draw[neutrino] (1,-1.3) -- (1,1.3);
 \draw[lepton] (2,-1.3) -- (1,-1.3);
\node at (0.5,-0.7) {$W$};
\node at (0.5,1.9) {$W$};
\node at (-0.25,-2.3) {(f)};
\end{tikzpicture}

\caption{Sample Feynman diagrams contributing to the $\gamma Z$ and to the $WW$ subsets.}
\label{fig:EWbosons}
\end{figure}

In the interference term  $\langle {\cal M}^{(0)} | {\cal M}^{(1,1)} \rangle$
we recognise different subsets associated to different combinations of the EW bosons exchanged in the loops.
We consider the following cases: $\gamma\gamma,\,\gamma Z,\,ZZ,\,W$;
in the neutral cases there are either one boson in a loop and one in a tree-level line
or both bosons in the loops, as depicted in a few representative cases
in Figure \ref{fig:EWbosons} a-c for the $\gamma Z$ subset;
in the charged case we find one or two $W$s always in the loop lines
Figure \ref{fig:EWbosons} d-f.
%%%%%%%%%%%%%%%%%%%%%%%%%%%%%%%%%%%%%%%%%%%%%%%%%%%%%%%%%%%%%%%%
The calculation of the bare matrix elements up to two-loop follows
a general procedure.
For the generation of the Feynman diagrams we have used two completely independent approaches,
one based on {\tt QGRAF}~\cite{Nogueira:1991ex} and a second one on {\tt FeynArts} \cite{Hahn:2000kx}.
Two independent in-house sets of routines,
written in {\tt FORM} \cite{Vermaseren:2000nd} and {\tt Mathematica} \cite{Mathematica}
have been used to perform the Lorentz and Dirac algebra.
After these first manipulations, the intermediate expressions contain
a large number of scalar Feynman integrals,
which can be reduced to a much smaller set of MIs by means of IBP and LI identities.
We have executed the reduction algorithms with two independent in-house programs based on the public codes
{\sc Kira} \cite{Maierhofer:2017gsa}, {\sc LiteRed}~\cite{Lee:2013mka, Lee:2012cn} and
{\sc Reduze}2~\cite{vonManteuffel:2012np, Studerus:2009ye}.
We have cross-checked the corresponding expressions at one- and two-loop level,
  obtained in the two independent approaches,
finding perfect agreement.

The evaluation of the quark wave function corrections
(Figure~\ref{fig:diagsamples}-b)
deserves a comment because it proceeds in two steps.
First, we generate all the Feynman diagrams contributing to the fermion self-energy
and reduce their Feynman integrals to MIs.
Second, since we have to compute the derivative of the self-energy with respect to the external invariant,
we express the derivative of the MIs as a combination of MIs with a standard reduction procedure.
We eventually evaluate this combination in the on-shell limit.

To simplify the calculation,
we consider an approximation of the scattering amplitude in the small lepton mass limit,
i.e. when $\ml$ is negligible compared
to the masses of the gauge bosons and to the energy scales of the process.
We thus consider the ratio $\ml/\sqrt{s}$
and keep only terms enhanced by a factor $\log(\ml^2/s)$,
while we discard all those contributions vanishing in the $\ml\to 0$ limit.
This approximation is used in the reduction to MIs procedure,
in order to distinguish the integrals where the dependence on $\ml$ has to be considered
from those where the massless lepton limit can be immediately taken,
with a remarkable simplification in the latter case.
After such procedure,
the only contributions with an explicit leftover dependence on the lepton mass
are the ones stemming from factorisable contributions with initial-state QCD and
one-loop final-state vertex corrections.
Among the box diagrams, those with one or two photons exchanges
individually develop logarithms of the lepton mass,
but the latter eventually cancel \cite{Frenkel:1976bj}
in the physical amplitude.

The identification of all the MIs which contribute to the final result
is given
using the notion of integral family, which we present in the next Section.

\subsection{Integral families and the representation of the result in terms of MIs}
We use the word topology to identify the flow of the external momenta
in the internal lines of a Feynman integral.
An integral family is defined as a complete set of denominator factors,
sufficient to reduce all the scalar products present in one scalar Feynman integral
in terms of linear combinations of these quantities.
In general, an integral family includes more denominator factors
than those appearing in a specific Feynman diagram.
In the present calculation we discuss two-loop four-point topologies
and we find that the corresponding integral families contain 9 denominator factors,
i.e. two more than the maximum number of internal lines of the Feynman diagram.
The Feynman integrals with a smaller number of internal lines
are called subtopologies, and belong to the same integral family of the
integral under study.

By associating  a scalar integral to a specific integral family,
we can express all the scalar products in
terms of the denominators of that particular family.
We thus obtain, after all the algebraic simplifications, integrals with the following form:
\begin{align}
I &= \int \frac{d^dk_1}{(2\pi)^d}\,  \frac{d^dk_2}{(2\pi)^d}\,
\frac{ 1 }{
  \left[ k_1^2 \right]^{\alpha_1}
  \left[ k_2^2 \right]^{\alpha_2}
  \left[ (k_1-k_2)^2-\mzsq  \right]^{\alpha_3}
  \left[ (k_2-p_1)^2 \right]^{\alpha_4}
  \cdots
}\nonumber\\
&\equiv
\int \frac{d^dk_1}{(2\pi)^d}\,  \frac{d^dk_2}{(2\pi)^d}\,
\frac{ 1 }{
{\cal D}_1^{\alpha_1}
{\cal D}_2^{\alpha_2}
[{\cal D}_{12} - \mzsq ]^{\alpha_3}
{\cal D}_{2;1}^{\alpha_4}
\cdots
  }\;,
\label{eq:exint}
\end{align}
where
the exponents $\alpha_k$ can be either positive (0, 1 or 2) or negative.
We introduce the symbols $\cD$s
\begin{equation}
  \cD_{i} = k_{i}^2, \cD_{ij} = (k_i-k_j)^2,
  \cD_{i;j} = (k_i-p_j)^2, \cD_{i;jl} = (k_i-p_j-p_l)^2, \cD_{ij;l} = (k_i-k_j-p_l)^2
\end{equation}
and we use them to precisely define the various integral families.
Those relevant for the full calculation are listed here below.

In the cases where the lepton masses are required,
because of the presence of final-state collinear singularities,
we find the following integral families which accommodate
all the appearing scalar integrals:
\begin{align}
  \label{eq:Basisml}
  {\rm B}_{0} &: \{ \cD_1, \cD_2, \cD_{12}, \cD_{1;1}, \cD_{2;1}, \cD_{1;12}, \cD_{2;12}, \cD_{1;3}, \cD_{2;3} \}
              \nonumber\\
  {\rm B}_{1} &: \{ \cD_1, \cD_2, \cD_{12}, \cD_{1;1}, \cD_{2;1}, \cD_{1;12}, \cD_{12;2}, \cD_{1;3}, \cD_{2;3} \}
              \nonumber\\
  {\rm B}_{20} &: \{ \cD_1, \cD_2, \cD_{12}, \cD_{1;1}, \cD_{2;1}, \cD_{1;12}, \cD_{2;12}, \cD_{1;3}-\ml^2, \cD_{2;3}-\ml^2 \}
              \nonumber\\
  {\rm B}_{20p} &: \{ \cD_1, \cD_2, \cD_{12}, \cD_{1;2}, \cD_{2;2}, \cD_{1;12}, \cD_{2;12}, \cD_{1;3}-\ml^2, \cD_{2;3}-\ml^2 \}
              \nonumber\\
  {\rm B}_{27} &: \{ \cD_1, \cD_2-\ml^2, \cD_{12}-\ml^2, \cD_{1;1}, \cD_{2;1}, \cD_{1;12}, \cD_{2;12}-\ml^2, \cD_{1;3}, \cD_{2;3} \}\, .
\end{align}
Additional families, where the lepton masses can be immediately discarded, are:
\begin{align}
  \label{eq:Basis0}
  {\rm B}_{11} &: \{ \cD_1, \cD_2, \cD_{12}-\mu_V^2, \cD_{1;1}, \cD_{2;1}, \cD_{1;12}, \cD_{2;12}, \cD_{1;3}, \cD_{2;3} \}
              \nonumber\\
  {\rm B}_{12} &: \{ \cD_1, \cD_2, \cD_{12}, \cD_{1;1}-\mu_V^2, \cD_{2;1}, \cD_{1;12}, \cD_{2;12}, \cD_{1;3}, \cD_{2;3} \}
              \nonumber\\
  {\rm B}_{13} &: \{ \cD_1, \cD_2, \cD_{12}-\mu_V^2, \cD_{1;1}, \cD_{2;1}, \cD_{1;12}, \cD_{12;2}, \cD_{1;3}, \cD_{2;3} \}
              \nonumber\\
  {\rm B}_{13p} &: \{ \cD_1, \cD_2, \cD_{12}-\mu_V^2, \cD_{1;2}, \cD_{2;2}, \cD_{1;12}, \cD_{12;1}, \cD_{1;3}, \cD_{2;3} \}
              \nonumber\\
  {\rm B}_{14} &: \{ \cD_1, \cD_2-\mu_V^2, \cD_{12}, \cD_{1;1}, \cD_{2;1}, \cD_{1;12}, \cD_{2;12}, \cD_{1;3}, \cD_{2;3} \}
              \nonumber\\
  {\rm B}_{14p} &: \{ \cD_1, \cD_2-\mu_V^2, \cD_{12}, \cD_{1;2}, \cD_{2;2}, \cD_{1;12}, \cD_{2;12}, \cD_{1;3}, \cD_{2;3} \}
              \nonumber\\
  {\rm B}_{16} &: \{ \cD_1, \cD_2-\mu_V^2, \cD_{12}, \cD_{1;1}, \cD_{2;1}, \cD_{1;12}, \cD_{2;12}-\mu_V^2, \cD_{1;3}, \cD_{2;3} \}
              \nonumber\\
  {\rm B}_{16p} &: \{ \cD_1, \cD_2-\mu_V^2, \cD_{12}, \cD_{1;2}, \cD_{2;2}, \cD_{1;12}, \cD_{2;12}-\mu_V^2, \cD_{1;3}, \cD_{2;3} \}
              \nonumber\\
  {\rm B}_{18} &: \{ \cD_1, \cD_2, \cD_{12}, \cD_{1;1}, \cD_{2;1}, \cD_{1;12}, \cD_{2;12}, \cD_{1;3}-\mu_V^2, \cD_{2;3}-\mu_V^2 \}\,,
\end{align}
where $V$ can be either $Z$ or $W$.
Each MI is thus uniquely identified by its integral family and the exponents $\alpha_j$,
e.g. $B_{20}[1,1,1,0,1,1,0,1,0 ]$ is the symbolic form of
\begin{equation}
 \int \frac{d^dk_1}{(2\pi)^d}\,  \frac{d^dk_2}{(2\pi)^d}\,
\frac{ 1 }{
   k_1^2    k_2^2   (k_1-k_2)^2
  (k_2-p_1)^2  (k_1-p_1-p_2)^2   ( (k_1-p_3)^2 -m_l^2) } \,.
\end{equation}

In the following, we list the appearing MIs for all the integral families.
The massless two-loop form factor integrals were presented in \cite{Gehrmann:2005pd}:
\begin{align}
B_{0} :&
\{0, 1, 1, 0, 0, 1, 0, 0, 0\}, \{0, 1, 1, 1, 0, 0, 1, 0, 0\}, \{1, 1, 0, 0, 0, 1, 1, 0, 0\}
\nonumber \\
B_{1} :&
\{1, 1, 1, 0, 1, 1, 1, 0, 0\}\;.
\label{eq:miB0B1}
\end{align}
The MIs with the massive lepton can be straightforwardly obtained from the planar MIs \cite{Bonciani:2008az, Bonciani:2009nb}
of NNLO QCD corrections to top
quark pair production by replacing the top quark mass with the lepton mass:
\begin{align}
B_{20} :&
\{0, 0, 1, 0, 0, 0, 1, 1, 0\}, \{0, 0, 1, 1, 0, 0, 1, 1, 0\}, \{0, 0, 2, 0, 1, 0, 0, 1, 0\},  \nn
\{0, 0, 1, 0, 1, 0, 0, 1, 0\}, \{0, 1, 1, 0, 0, 1, 0, 1, 0\}, \{0, 1, 1, 0, 1, 1, 0, 1, 0\},  \nn
\{0, 1, 1, 1, 0, 1, 0, 1, 0\}, \{0, 1, 2, 0, 1, 1, 0, 1, 0\}, \{1, 0, 1, 0, 1, 1, 0, 1, 0\},  \nn
\{1, 0, 2, 0, 1, 1, 0, 1, 0\}, \{0, 1, 0, 0, 0, 0, 1, 1, 0\}, \{0, 1, 1, 0, 0, 0, 1, 1, 0\},  \nn
\{0, 1, 1, 0, 1, 0, 1, 1, 0\}, \{0, 2, 1, 0, 0, 0, 1, 1, 0\}, \{0, 1, 2, 0, 0, 0, 1, 1, 0\},  \nn
\{0, 2, 1, 0, 1, 0, 1, 1, 0\}, \{1, 1, 0, 0, 0, 1, 1, 1, 0\}, \{1, 1, 1, 0, 1, 1, 1, 1, 0\},  \nn
\{1, 1, 1, 0, 1, 1, 1, 2, 0\}, \{1, 1, 2, 0, 1, 1, 1, 1, 0\}
\nonumber \\
B_{20p} :&
\{0, 0, 1, 1, 0, 0, 1, 1, 0\}, \{0, 0, 2, 0, 1, 0, 0, 1, 0\}, \{0, 0, 1, 0, 1, 0, 0, 1, 0\},  \nn
\{0, 1, 1, 0, 1, 1, 0, 1, 0\}, \{0, 1, 1, 1, 0, 1, 0, 1, 0\}, \{0, 1, 2, 0, 1, 1, 0, 1, 0\},  \nn
\{1, 0, 1, 0, 1, 1, 0, 1, 0\}, \{1, 0, 2, 0, 1, 1, 0, 1, 0\}, \{0, 1, 1, 0, 1, 0, 1, 1, 0\},  \nn
\{0, 2, 1, 0, 1, 0, 1, 1, 0\}, \{1, 1, 1, 0, 1, 1, 1, 1, 0\}, \{1, 1, 1, 0, 1, 1, 1, 2, 0\},  \nn
\{1, 1, 2, 0, 1, 1, 1, 1, 0\}
\nonumber\\
B_{27} :& \{1, 1, 0, 0, 0, 1, 1, 0, 0\}\;.
\label{eq:miB20}
\end{align}
The form factor type two-loop MIs with one or two massive lines were obtained in \cite{Aglietti:2003yc,Aglietti:2004tq}:
\begin{align}
B_{11} :&
\{0, 0, 1, 0, 0, 1, 1, 0, 0\}, \{0, 1, 1, 0, 0, 0, 1, 0, 0\}, \{0, 1, 1, 0, 0, 1, 0, 0, 0\},  \nn
\{0, 2, 1, 0, 0, 1, 0, 0, 0\}, \{0, 1, 1, 1, 0, 0, 1, 0, 0\}, \{0, 2, 1, 1, 0, 0, 1, 0, 0\},  \nn
\{0, 1, 1, 1, 0, 1, 1, 0, 0\}, \{1, 1, 1, 0, 0, 1, 1, 0, 0\}
\nonumber\\
B_{12} :&
\{0, 1, 1, 1, 0, 1, 0, 0, 0\}, \{0, 1, 1, 1, 0, 1, 1, 0, 0\}, \{1, 0, 1, 1, 1, 1, 0, 0, 0\},  \nn
\{1, 1, 0, 1, 0, 1, 1, 0, 0\}
\nonumber\\
B_{13} :&
\{1, 1, 1, 0, 0, 1, 1, 0, 0\}, \{2, 1, 1, 0, 0, 1, 1, 0, 0\}, \{1, 1, 1, 0, 1, 1, 1, 0, 0\}\;.
\label{eq:miB11}
\end{align}
The two-loop box MIs with one and two massive lines have been computed in \cite{Bonciani:2016ypc,Heller:2019gkq,Hasan:2020vwn}.
We relate our MIs as given below to the ones obtained in \cite{Bonciani:2016ypc} through the IBP reduction
and obtain the results for our MIs:
\begin{align}
B_{14} :&
\{0, 0, 1, 0, 1, 1, 0, 0, 1\}, \{0, 0, 1, 1, 0, 0, 0, 0, 1\}, \{0, 1, 1, 0, 1, 1, 0, 0, 1\},  \nn
\{0, 1, 1, 1, 0, 0, 0, 0, 1\}, \{0, 1, 1, 1, 0, 0, 1, 0, 0\}, \{0, 1, 1, 1, 0, 0, 1, 0, 1\},  \nn
\{0, 1, 1, 1, 0, 1, 0, 0, 1\}, \{0, 2, 1, 0, 1, 1, 0, 0, 1\}, \{0, 2, 1, 1, 0, 1, 0, 0, 1\},  \nn
\{1, 0, 1, 0, 1, 0, 1, 0, 1\}, \{1, 0, 1, 1, 0, 0, 1, 0, 1\}, \{1, 1, 1, 0, 0, 0, 1, 0, 0\},  \nn
\{1, 1, 1, 0, 1, 0, 0, 0, 1\}, \{1, 1, 1, 0, 1, 0, 1, 0, 1\}, \{1, 1, 1, 1, 0, 0, 0, 0, 1\},  \nn
\{1, 1, 1, 1, 0, 0, 1, 0, 0\}, \{1, 1, 1, 1, 0, 0, 1, 0, 1\}, \{1, 0, 1, 1, 0, 1, 0, 0, 1\},  \nn
\{1, 1, 0, 0, 0, 1, 1, 0, 0\}, \{1, 1, 1, 1, 0, 1, 0, 0, 1\}, \{1, 1, 1, 1, 0, 1, 1, 0, 1\},  \nn
\{2, 1, 1, 1, 0, 1, 0, 0, 1\}, \{1, 1, 2, 1, 0, 1, 1, 0, 1\}
\nonumber\\
B_{14p} :&
\{0, 0, 1, 0, 1, 1, 0, 0, 1\}, \{0, 0, 1, 1, 0, 0, 0, 0, 1\}, \{0, 1, 1, 0, 1, 1, 0, 0, 1\},  \nn
\{0, 1, 1, 1, 0, 0, 0, 0, 1\}, \{0, 1, 1, 1, 0, 0, 1, 0, 1\}, \{0, 1, 1, 1, 0, 1, 0, 0, 1\},  \nn
\{0, 2, 1, 0, 1, 1, 0, 0, 1\}, \{0, 2, 1, 1, 0, 1, 0, 0, 1\}, \{1, 0, 1, 0, 1, 0, 1, 0, 1\},  \nn
\{1, 0, 1, 1, 0, 0, 1, 0, 1\}, \{1, 1, 1, 0, 1, 0, 0, 0, 1\}, \{1, 1, 1, 0, 1, 0, 1, 0, 1\},  \nn
\{1, 1, 1, 1, 0, 0, 0, 0, 1\}, \{1, 1, 1, 1, 0, 0, 1, 0, 1\}, \{1, 0, 1, 1, 0, 1, 0, 0, 1\},  \nn
\{1, 1, 1, 1, 0, 1, 0, 0, 1\}, \{1, 1, 1, 1, 0, 1, 1, 0, 1\}, \{1, 1, 2, 1, 0, 1, 1, 0, 1\},  \nn
\{2, 1, 1, 1, 0, 1, 0, 0, 1\}
\nonumber\\
B_{16} :&
\{0, 1, 1, 0, 0, 1, 1, 0, 0\}, \{0, 1, 1, 0, 0, 1, 1, 0, 1\}, \{0, 1, 1, 0, 1, 1, 1, 0, 1\},  \nn
\{0, 1, 1, 1, 0, 0, 1, 0, 0\}, \{0, 1, 1, 1, 0, 0, 1, 0, 1\}, \{0, 1, 1, 1, 0, 1, 1, 0, 0\},  \nn
\{0, 1, 1, 1, 0, 1, 1, 0, 1\}, \{0, 2, 1, 1, 0, 0, 1, 0, 0\}, \{0, 2, 1, 1, 0, 0, 1, 0, 1\},  \nn
\{1, 1, 0, 0, 0, 1, 1, 0, 0\}, \{1, 1, 0, 0, 0, 1, 1, 0, 1\}, \{1, 1, 1, 0, 0, 1, 1, 0, 0\},  \nn
\{1, 1, 1, 0, 0, 1, 1, 0, 1\},
%{\color{blue}
%\{1, 1, 1, 1, 0, 1, 1, 0, 1\}, \{1, 1, 1, 2, 0, 1, 1, 0, 1\},  \nn
%\{1, 2, 1, 1, 0, 1, 1, 0, 1\}, \{2, 1, 1, 1, 0, 1, 1, 0, 1\}   \nn
%}
  \{1, 1, 1, 1, 0, 1, 1, 0, 1\},  \{1, 1, 1, 1, -1, 1, 1, 0, 1\},  \nn
  \{1, 1, 1, 1, 0, 1, 1, -1, 1\}, \{1, 1, 1, 1, -1, 1, 1, -1, 1\}
\nonumber\\
B_{16p} :&
\{0, 1, 1, 0, 1, 1, 1, 0, 1\}, \{0, 1, 1, 1, 0, 0, 1, 0, 1\}, \{0, 1, 1, 1, 0, 1, 1, 0, 1\},  \nn
\{0, 2, 1, 1, 0, 0, 1, 0, 1\},
%{\color{blue}
%\{1, 1, 1, 1, 0, 1, 1, 0, 1\}, \{1, 1, 1, 2, 0, 1, 1, 0, 1\},  \nn
%\{1, 2, 1, 1, 0, 1, 1, 0, 1\}, \{2, 1, 1, 1, 0, 1, 1, 0, 1\}   \nn
%}
  \{1, 1, 1, 1, 0, 1, 1, 0, 1\},  \{1, 1, 1, 1, -1, 1, 1, 0, 1\},  \nn
  \{1, 1, 1, 1, 0, 1, 1, -1, 1\}, \{1, 1, 1, 1, -1, 1, 1, -1, 1\}
\nonumber\\
B_{18} :& \{1, 1, 0, 0, 0, 1, 1, 1, 0\}\;.
\label{eq:miB14}
\end{align}
The reduction to the MIs allows us to obtain an intermediate representation
of the bare results, written as a linear combination of MIs with the corresponding rational coefficients
\begin{equation}
\langle \unM^{(0)} | \unM^{(1,1)} \rangle \, = \sum_{k=1}^{204} c_k(s,t,\{m_i\},\varepsilon) I_k(s,t,\{m_i\},\varepsilon)
\label{eq:sumint}
\end{equation}
with both the coefficients $c_k$ and the MIs $I_k$ being dependent on the dimensional regularisation parameter $\varepsilon$,
the kinematical variables $s,t$ and the masses $\{m_i\}$ of the gauge bosons and fermions.
The integrals $I_k$ are all of the form $B_i[\vec\alpha_j]$,
where $i$ labels
the integral family and $\vec\alpha_j$ is the set of all the exponents.
Due to their large size, we provide the explicit expressions of the results in the attached ancillary files.
In \textsf{bareNCDY.m}, we present the unrenormalised matrix elements
of the vertex and box corrections (Figures~\ref{fig:diagsamples}-a, \ref{fig:diagsamples}-f),
in terms of the MIs with the corresponding rational coefficients schematically described in eq.~\eqref{eq:sumint},
as a useful intermediate representation.

\section{Infrared Singularities and Universal Pole Structure}
\label{sec:infra}
The renormalised matrix elements are UV-finite,
but still contain IR divergences
originating from soft and/or collinear massless partons.
These singularities are guaranteed to cancel
when combined with the NNLO QCD-EW real and real-virtual corrections
and subtracted of the initial-state collinear poles,
to obtain an IR-safe observable.
Nevertheless,
their presence at intermediate stages of an inclusive computation,
as it is the case with the results provided in this paper,
requires the development of methods to consistently handle them;
they must be systematically subtracted, leaving a finite remnant.

While at NLO computations the Catani-Seymour dipole subtraction~\cite{Catani:1996jh,Catani:1996vz,Catani:2002hc} and FKS subtraction~\cite{Frixione:1995ms} are the two methods most widely used, at NNLO several techniques have been proposed (see e.g. \cite{Proceedings:2018jsb} and references therein), each one with a different subtraction procedure for the various contributions to physical observables.

The subtraction of the IR poles from the two-loop matrix elements is in general achieved via a process-independent subtraction operator $\mathcal{I}$,
that can be constructed by using the universality of the
IR singularity structure of the scattering amplitudes,
known in the case of a massless gauge theory up to two-loop level
\cite{Catani:1998bh,Sterman:2002qn,Becher:2009cu,Gardi:2009qi}.
An explicit study was performed in \cite{Kilgore:2011pa,Kilgore:2013uta}
to obtain the IR structure of the two-loop amplitudes
for mixed QCD$\otimes$QED corrections to NC Drell-Yan production
considering massless leptons.
The case of theories with massive particles  has been studied in
\cite{Mitov:2006xs,Becher:2007cu,Becher:2009kw,Ahmed:2017gyt,Blumlein:2018tmz}.
In particular, the IR structure for one-loop QCD corrections
to top quark pair production has been examined in detail in ref.\cite{Catani:2014qha}, while at two-loop level in refs.\cite{Catani:2019iny,Catani:2019hip,Catani:2020kkl}.
This can be appropriately abelianised \cite{Buonocore:2019puv} to obtain
the IR structure in the present case
by replacing the top quarks with massive leptons.

Different subtraction operator within different subtraction schemes can in principle differ from each other:
anyhow the universality of the IR divergences implies that the subtracted virtual contribution at two loops may differ, at most, by a finite constant.
For the sake of definiteness,
we illustrate here the subtraction operator used in our calculation, which has been performed
in the framework of the $q_T$-subtraction formalism~\cite{Catani:2007vq}.
However, we stress that our results can be easily converted and combined to the
description of real radiation in any other IR subtraction scheme,
by adding the appropriate finite remainder.

The IR subtraction functions ($\I$) at one-loop are given by
\begin{align}
\I^{(1,0)} &= \smu^{-\ep} C_F \left( - \frac{2}{\ep^2} - \frac{1}{\ep} (3 + 2 i \pi)  + \zeta_2 \right) \,,
\\
\I^{(0,1)} &= \smu^{-\ep} \bigg[ Q_u^2 \left( - \frac{2}{\ep^2} - \frac{1}{\ep} (3 + 2 i \pi)  + \zeta_2 \right)
 + \frac{4}{\ep} {\Gamma}_{l}^{(0,1)} \bigg] \,,
\end{align}
where
\begin{equation}
 {\Gamma}_{l}^{(0,1)} = Q_u Q_l \log \bigg( \frac{2 p_1.p_3}{2 p_2.p_3} \bigg)
 + \frac{Q_l^2}{2} \bigg(  -1 - \frac{1+x_l^2}{1-x_l^2} \log (x_l) \bigg) \,.
\end{equation}
$Q_l$ and $Q_u$ are the charges of the lepton and of the initial-state quark, and the Casimir of the fundamental representation of SU(N), $C_F$, is given by $C_F=\frac{N^2-1}{2 N}$ \,.
The variable $x_l$ is defined by
\be
 \frac{(1-x_l)^2}{x_l} = -\frac{s}{\ml^2}\;.
\label{eq:defxl}
\ee
Using the one-loop subtraction functions, we obtain the finite contributions to
the one-loop QCD and EW amplitudes, respectively, as follows:
\begin{align}
 | \M^{(1,0),fin} \rangle &= | \M^{(1,0)} \rangle -  \I^{(1,0)} | \M^{(0)} \rangle \,,
 \label{eq:QCDfin}\\
 | \M^{(0,1),fin} \rangle &= | \M^{(0,1)} \rangle -  \I^{(0,1)} | \M^{(0)} \rangle \,.
 \label{eq:EWfin}
\end{align}
The mixed two-loop subtraction operator\footnote{
An analogous expression, in the massless lepton case, has been discussed in refs.~\cite{Kilgore:2011pa,Kilgore:2013uta}.
  }
is given by
%%%%%%%%%%%%%%%%%%%%%%
%
\begin{align}
  \I^{(1,1)} = \smu^{-2\ep} C_F &
\bigg[
Q_u^2  \bigg( \frac{4}{\ep^4} + \frac{1}{\ep^3} ( 12 + 8 i \pi ) + \frac{1}{\ep^2} ( 9 - 28 \zeta_2 + 12 i \pi)
+ \frac{1}{\ep} \Big( -\frac{3}{2} + 6 \zeta_2
\nonumber\\&
- 24 \zeta_3 - 4 i \pi \zeta_2 \Big) \bigg)
+ \left( - \frac{2}{\ep^2} - \frac{1}{\ep} (3 + 2 i \pi)  + \zeta_2 \right)~\frac{4}{\ep}~\Gamma_l^{(0,1)} \bigg].
\label{eq:i11}
\end{align}
%% %
  In order to describe the contributions entering in $\I^{(1,1)}$,
  let us discuss first the IR singularities of the one-loop and two-loop boxes.
  The fact that the $q\bar q$ initial state has zero electric charge
  implies that there are no QED collinear singularities
  stemming from the exchange of a photon between an initial quark and a
  final lepton lines. Such terms appear in the individual diagrams, but cancel
  in the physical amplitude. This conclusion can be also seen as a confirmation
  of the universality of initial state QED collinear singularities.
  This property is observed at one loop and confirmed also when going to \oaas
  by dressing the initial state quark line with all the possible
  gluon exchanges. In the sum of all the two-loop box diagrams
  we can thus expect that the
  photon exchange develops at most a soft divergence.
  The presence of a colorful initial state quark line
  but of a colorless final state lepton line
  leads to a natural separation between the gluon IR-divergent
  exchanges in the initial state
  and the QED interaction in the rest of the diagram.
  For this reason, all the IR singularities,
  soft and collinear, stemming from the two-loop box diagrams, can be subtracted
  by the product of $\I^{(1,0)}$ and $\I^{(0,1)}$, times the Born amplitude.
  The divergences in the product of initial and final factorisable contributions
  are obviously canceled by the same product of one-loop subtraction operators.
  The initial state vertex corrections have a richer structure,
  including planar and non-planar topologies,
  both yielding IR divergent contributions.
  The former can be subtracted again by the product of
  $\I^{(1,0)}$ and $\I^{(0,1)}$,
  while the latter require a genuinely two-loop subtraction operator.
  From the above comments we conclude that the operator $\I^{(1,1)}$
  is applied to the Born amplitude and
  receives contributions from the product of $\I^{(1,0)}$ and $\I^{(0,1)}$
  and from the two-loop non-factorisable contributions appearing in the
  vertex corrections.

  We note that the application of $\I^{(1,1)}$ can not make the
  complete amplitude IR finite. There are in fact additional configurations
  where e.g. the gluon yields a divergence and the EW interaction is finite.
  This is for instance the case in the two-loop boxes with the exchange
  of two $W$ or $Z$ bosons.
  They can be subtracted by applying the one-loop QCD subtraction
  operator  $\I^{(1,0)}$ to the one-loop EW amplitude subtracted of its one-loop
  QED divergences, as defined in eq.~(\ref{eq:EWfin}).
  An analogous subtraction can remove the photon divergences which multiply
  a finite QCD remnant, as defined in eq.~(\ref{eq:QCDfin}).

Using eqs.~(\ref{eq:QCDfin}-\ref{eq:i11}), we obtain the finite and subtracted two-loop amplitude as follows
\begin{equation}
  | \M^{(1,1),fin} \rangle =
  | \M^{(1,1)} \rangle -  \I^{(1,1)} | \M^{(0)} \rangle
                      - \tilde{\I}^{(0,1)} | \M^{(1,0),fin} \rangle
                      - \tilde{\I}^{(1,0)} | \M^{(0,1),fin} \rangle \, ,
 \label{eq:subtracted}
\end{equation}
with $\tilde{\I}^{(i,j)}$  defined by dropping
the term proportional to $\zeta_2$ in ${\I}^{(i,j)}$.
This choice is conventional and defines the finite part of our subtraction term.

The approximation of the amplitude in the small lepton mass limit
retains all the terms enhanced by $\log (m_l)$,
divergent in the $\ml\to 0$ limit.
The structure of these corrections reflects the universality property
of the final-state collinear divergences,
and is given, normalised to the Born squared matrix element, by
\begin{equation}
  \lim_{\ml\to0}
  \frac{\langle \M^{(0)} | \M^{(1,1)} \rangle}{\langle \M^{(0)} | \M^{(0)}\rangle}=
K~+~
 C_F Q_l^2 ( -8+7 \zeta_2 - 3 i \pi )
\bigg[ - \log \bigg( \frac{\ml^2}{s} \bigg)
       + \log^2  \bigg( \frac{\ml^2}{s} \bigg)
\bigg] \,,
\end{equation}
%% %
where $K$ represents all the other terms in the interference,
constant in the $\ml\to 0$ limit.

\section{Results}
\label{sec:results}
\subsection{The UV-renormalised IR-subtracted hard function}
Using eq.~\eqref{eq:subtracted},
we obtain the finite IR-subtracted matrix element $\langle \M^{(0)} | \M^{(1,1),fin} \rangle$
from the UV renormalised matrix element $\langle \M^{(0)} | \M^{(1,1)} \rangle$
\begin{align}
\langle \M^{(0)} | \M^{(1,1),fin} \rangle &= \langle \M^{(0)} | \M^{(1,1)} \rangle -  \I^{(1,1)} \langle \M^{(0)}  | \M^{(0)} \rangle
\nonumber\\&
                                                 - \tilde{\I}^{(0,1)} \langle \M^{(0)} | \M^{(1,0),fin} \rangle
                                                 - \tilde{\I}^{(1,0)} \langle \M^{(0)} | \M^{(0,1),fin} \rangle \,.
\end{align}
  This quantity, renormalised with the Born squared matrix element, represents the hard function $H^{(1,1)}$,
  defined as
\begin{equation}
 H^{(1,1)} =
 \frac{1}{16}~
\left[
  2 ~ {\rm Re} \left( \frac{\langle \M^{(0)} | \M^{(1,1),fin} \rangle}{\langle \M^{(0)} | \M^{(0)} \rangle} \right)
\right]
  \;.
\end{equation}
The contributions to the two-loop vertex and box diagrams,
introduced in eq.~(\ref{eq:sumint}),
come from 204 MIs, listed in eqs.~(\ref{eq:miB0B1}-\ref{eq:miB14}).
We compute them
using the method of differential equations \cite{Kotikov:1990kg,Remiddi:1997ny,Gehrmann:1999as,Argeri:2007up,Henn:2014qga,Ablinger:2015tua,Ablinger:2018zwz}.
For a large subset, we express the solution
in terms of generalised harmonic polylogarithms\footnote{
We present in Appendix \ref{app:readme} the relevant kinematical variables, which appear in these GHPLs.}
(GHPLs) \cite{Goncharov:polylog,Goncharov:2001iea,Remiddi:1999ew}.
In our results we also find
the following 5 MIs
\begin{align}
  & \{0, 1, 1, 1, 0, 1, 1, 0, 1\},
  \{1, 1, 1, 1, 0, 1, 1, 0, 1\},  \{1, 1, 1, 1, -1, 1, 1, 0, 1\},  \nn
  \{1, 1, 1, 1, 0, 1, 1, -1, 1\}, \{1, 1, 1, 1, -1, 1, 1, -1, 1\}\, ,
\label{eq:boxmis}
\end{align}
each in the two integral families $B_{16}$ and $B_{16p}$
which are not expressed,
according to the discussion of ref.~\cite{Bonciani:2016ypc},
in terms of GHPLs,
but using Chen-Goncharov integrals \cite{Chen:1977oja}.
The possibility to solve the MIs with two internal massive lines
and obtain a representation in terms of Nielsen polylogarithmic functions
has been discussed in ref.~\cite{Heller:2019gkq}.
In our case, the evaluation of the Chen iterated integrals \cite{Chen:1977oja}
in a non-physical region, where the boundary conditions are fixed, is possible,
but the analytic continuation to cover the full physical region
constitutes a formidable task.
For this reason, we adopt a different, semi-analytical, approach
to solve and evaluate these MIs,
and we describe it in Section \ref{sec:semianalytical}.

The final results, expanded in powers of $\ep$,
are written in terms of GHPLs and of the symbols associated to
the 5 MIs whose solution is given in terms of Chen integrals.
In the next Sections,
we explicitly discuss the numerical evaluation of the latter.
We note that the individual contributions from these 5 MIs,
to the single pole in $\ep$ of the full unrenormalised matrix element,
contain the Chen iterated integrals,
but the summed result is independent of those.

\subsection{Semi-analytical solution of MIs via series expansion}
\label{sec:semianalytical}
\begin{figure}[!ht]
\begin{center}
\includegraphics[width=0.33\textwidth]{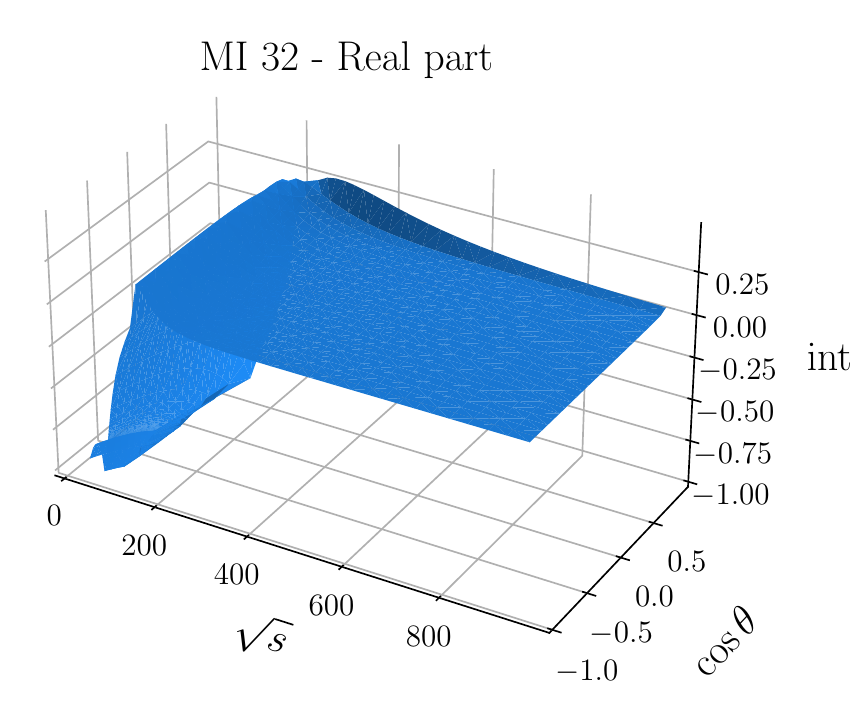}\hspace{2cm}
\includegraphics[width=0.33\textwidth]{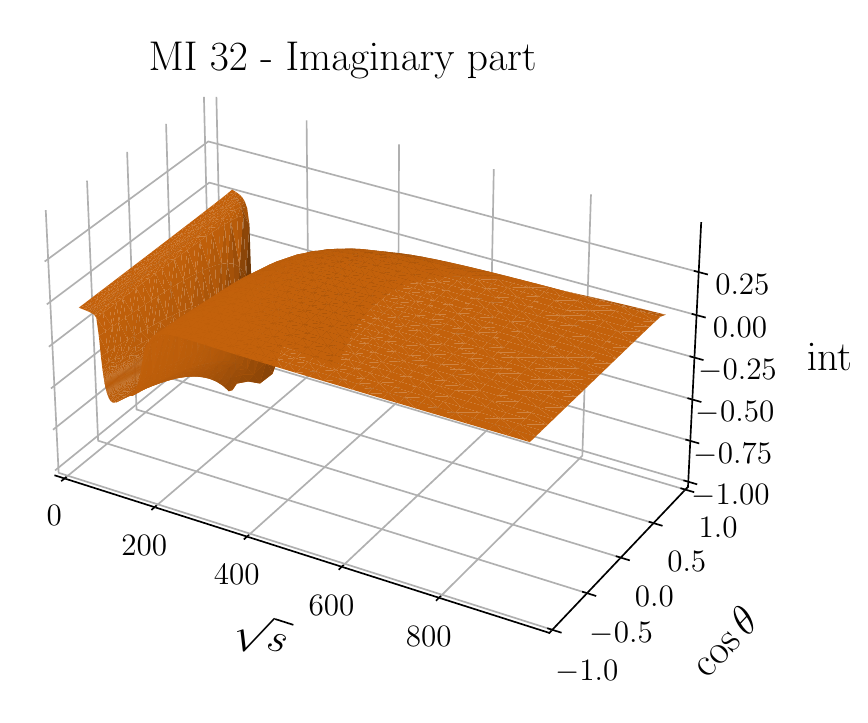}\\\vspace{-0.1cm}
\includegraphics[width=0.33\textwidth]{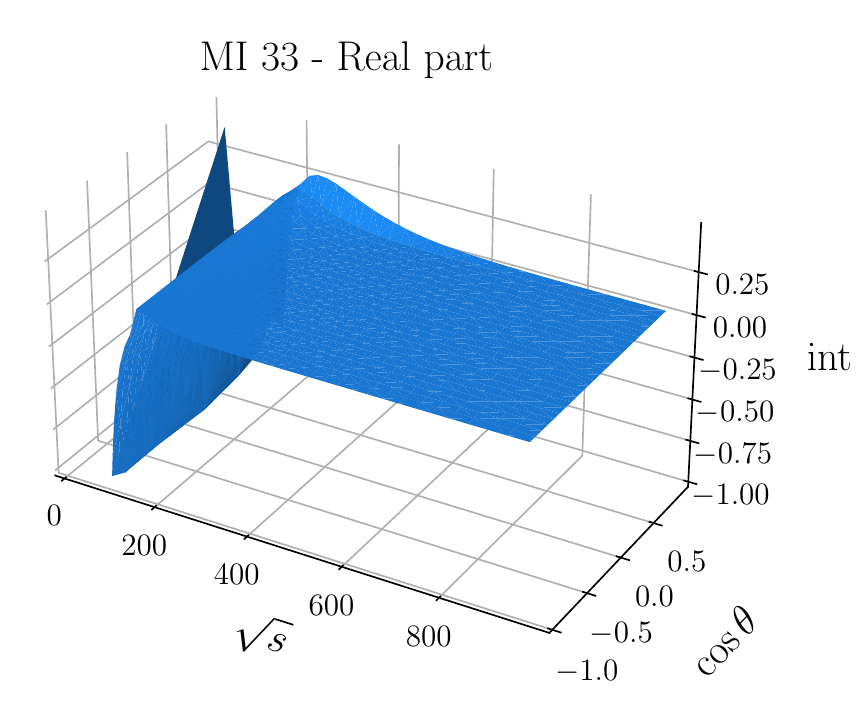}\hspace{2cm}
\includegraphics[width=0.33\textwidth]{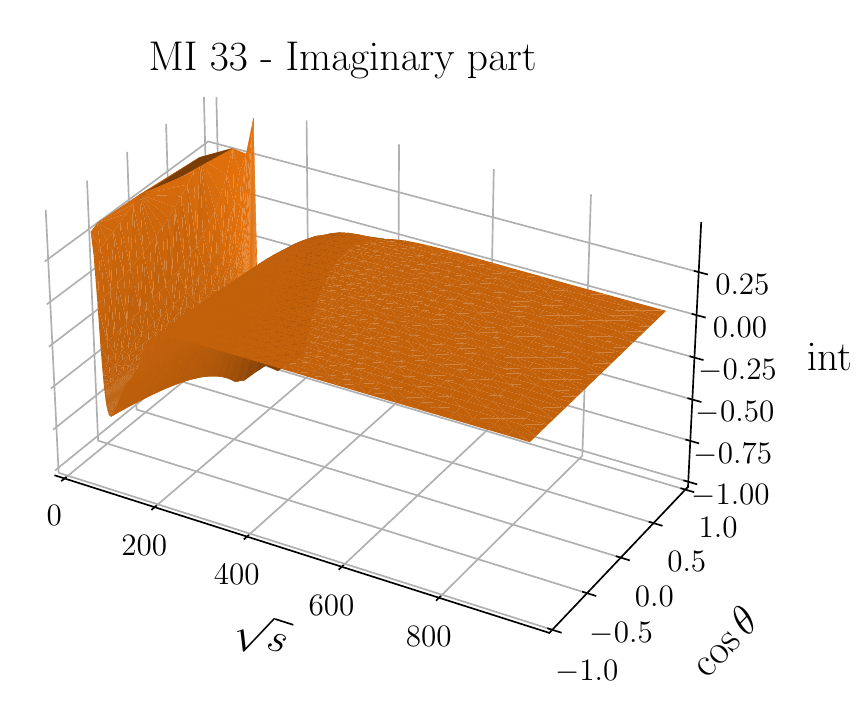}\\\vspace{-0.1cm}
\includegraphics[width=0.33\textwidth]{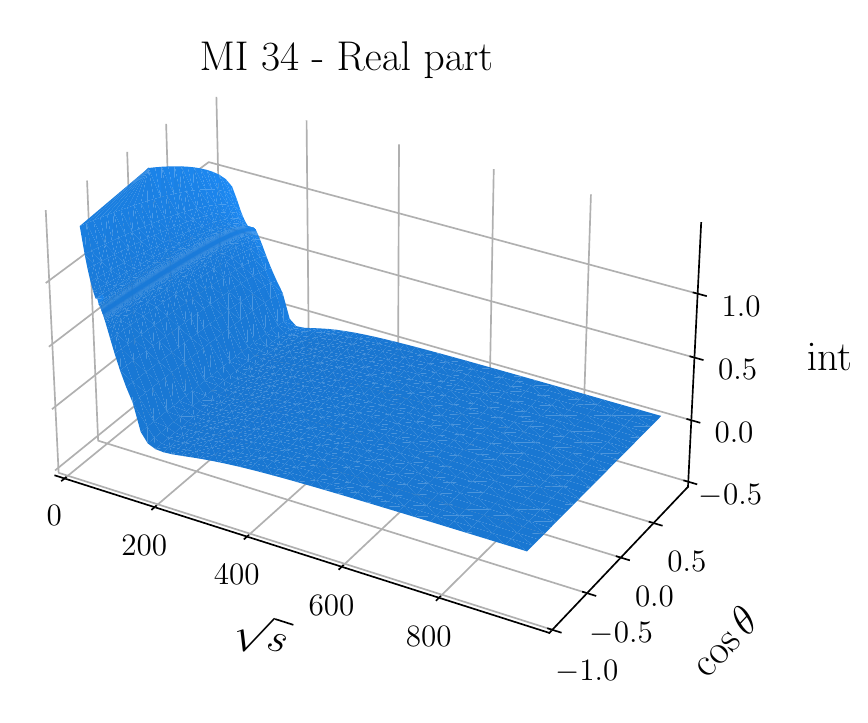}\hspace{2cm}
\includegraphics[width=0.33\textwidth]{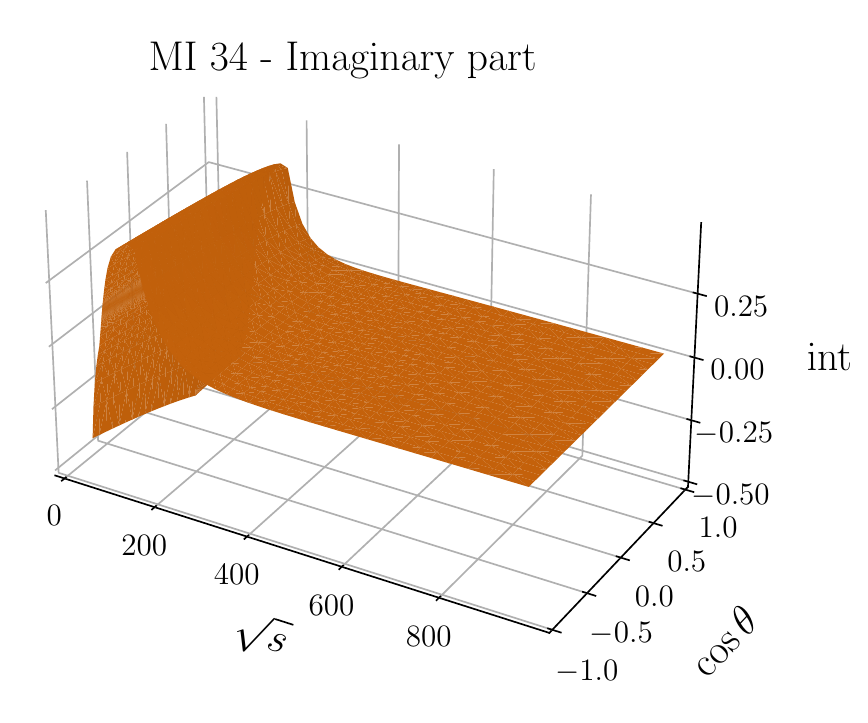}\\\vspace{-0.1cm}
\includegraphics[width=0.33\textwidth]{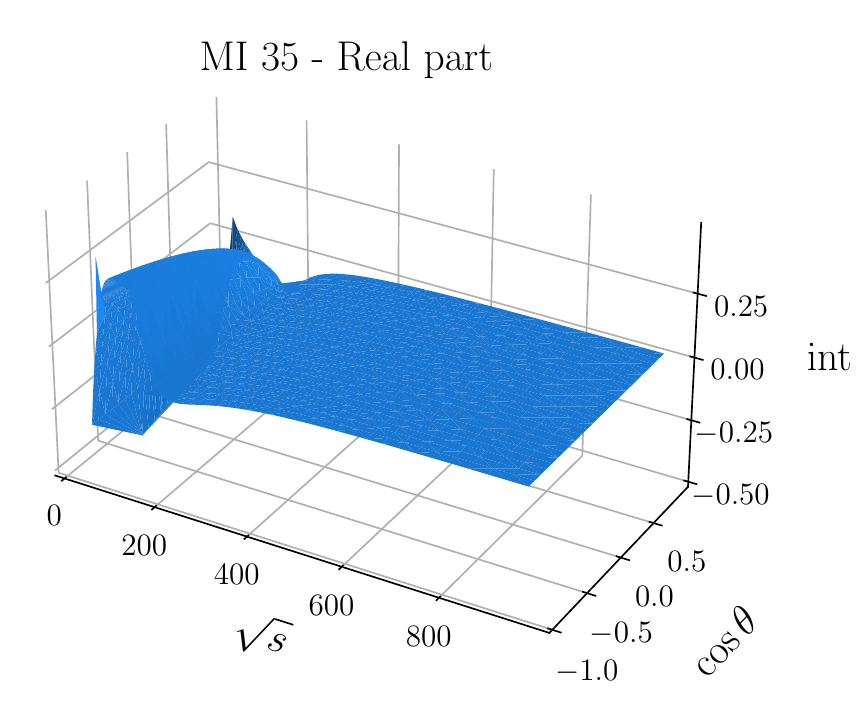}\hspace{2cm}
\includegraphics[width=0.33\textwidth]{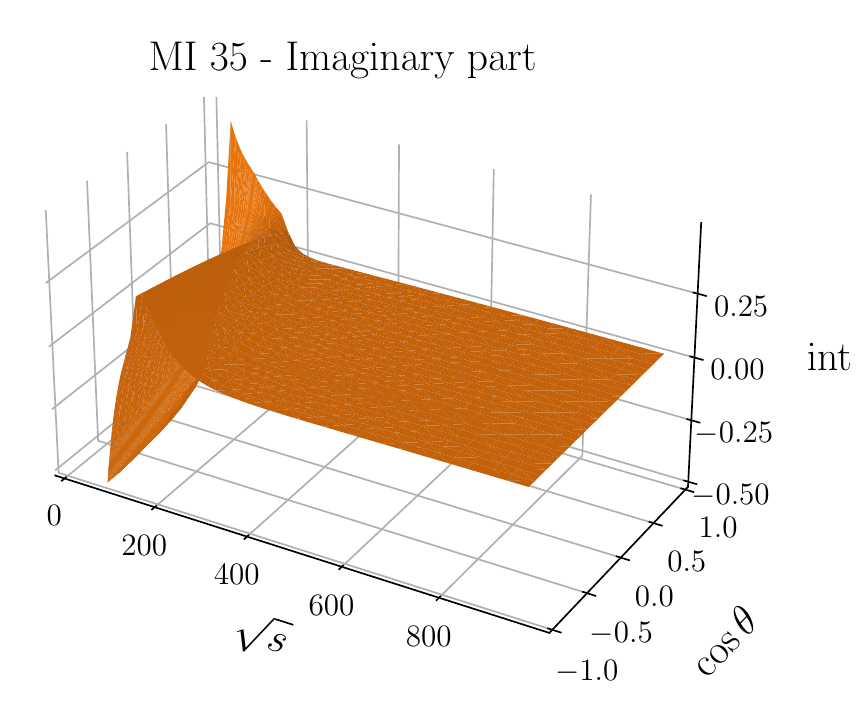}\\\vspace{-0.1cm}
\includegraphics[width=0.33\textwidth]{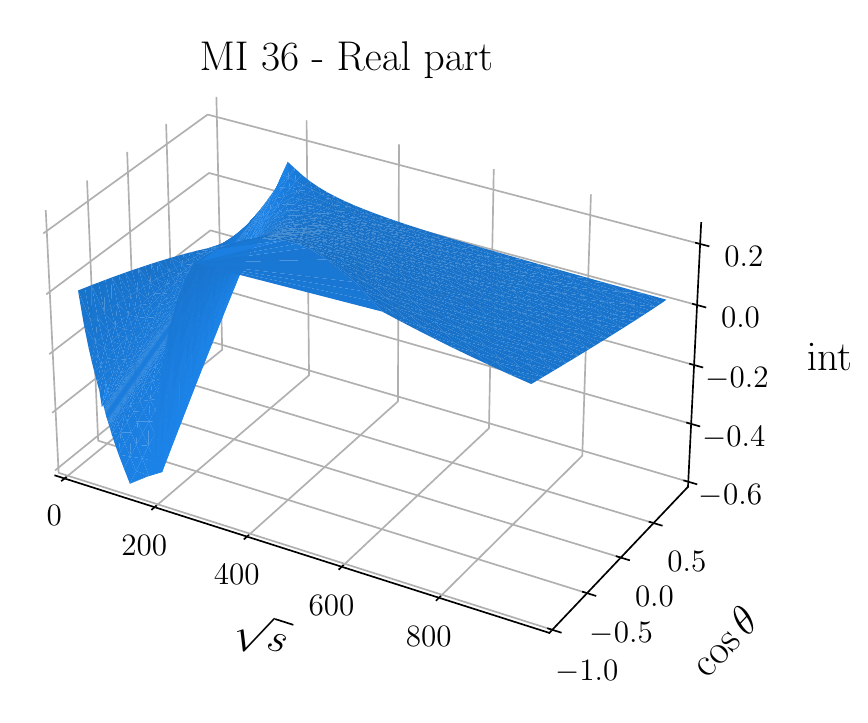}\hspace{2cm}
\includegraphics[width=0.33\textwidth]{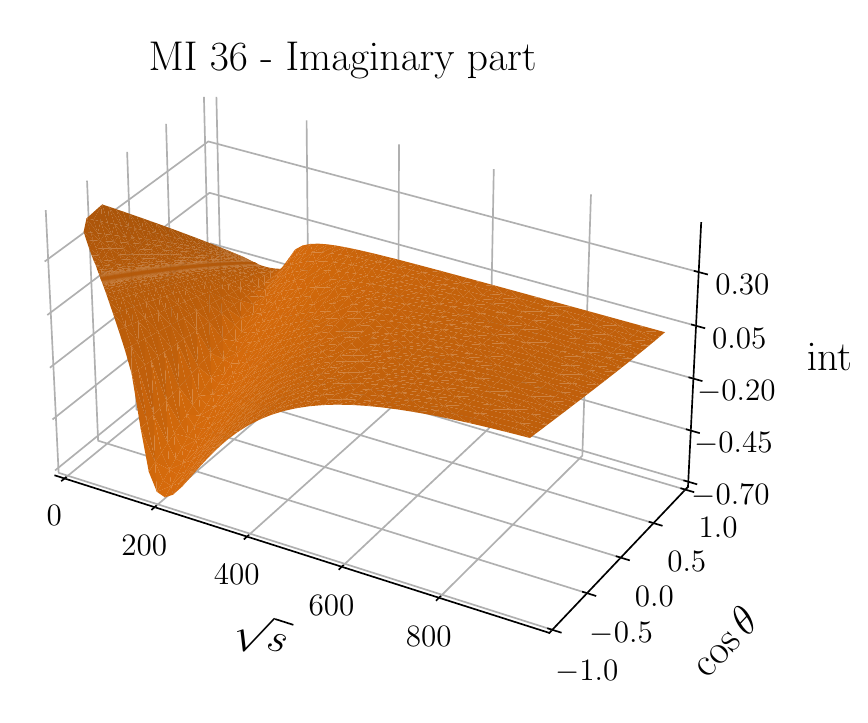}
\end{center}
\vspace{-1cm}
\caption{\label{fig:MIs}
Real (left column) and imaginary (right column) parts of the $\ep^0$ coefficient of the two-mass MIs labeled 32 to 36 in ref.~\cite{Bonciani:2016ypc}.
  }
\end{figure}

In ref.~\cite{Bonciani:2016ypc}, 5 MIs with two internal massive lines
have been solved in terms of Chen iterated integrals,
but their analytical continuation to
the physical region is extremely difficult.
We thus turn to a different approach and we obtain these MIs in semi-analytical form,
by solving the relevant system of differential equations via series expansion as initially proposed in ref.~\cite{Moriello:2019yhu}.
This approach is implemented in the {\tt Mathematica} code {\tt DiffExp} \cite{Hidding:2020ytt}.
However, since the public version of {\tt DiffExp} does not support, for the time being, the possibility of complex masses in the internal lines of the loop integrals, we have written an independent implementation of this procedure,
that consistently takes into account the presence of complex weights and the analytic continuation to the complex plane \cite{SeaFire}.

By approaching the problem in such semi-analytical way,
we have to consider the complete system ($36 \times 36$)
of differential equations \cite{Bonciani:2016ypc},
which includes the 5 MIs of  eq.~(\ref{eq:boxmis}).
The other 31 integrals are already known in terms of GHPLs
and we use their numerical evaluation obtained
with {\tt GiNaC} \cite{Vollinga:2004sn}
as a cross check of the correct behaviour of our code 
finding perfect agreement\footnote{
  Both codes work with arbitrary precision and
  any level of agreement can be reached simply adding more terms in the respective expansions.
In particular we have verified explicitly agreement of the first 40 significant digits }.
The values for the 5 MIs in eq.~(\ref{eq:boxmis}) are instead new predictions.
  In the euclidean region, we have checked our results for these 5 MIs against the values
  computed with {\tt PySecDec} \cite{Borowka:2017idc},
  with perfect agreement within the numerical precision of {\tt PySecDec}.
  In the physical region such comparison can not be performed,
  because of convergence issues observed with {\tt PySecDec},
  and we have adopted a different strategy.  
We have solved the same system of differential equations
with a real value for the gauge-boson masses,
by using both {\tt DiffExp} and our private implementation,
finding perfect agreement\footnote{
Also in this case the two codes can work with arbitrary precision and we have explicitly checked the agreement of the first 40 digits.},
  for different values of the centre-of-mass energy, below and above all the physical thresholds.

In Figure \ref{fig:MIs} we present the real and imaginary parts
of the values of the $\ep^0$ coefficient of the MIs numbered from 32 to 36 in ref.~\cite{Bonciani:2016ypc},
in a phase-space region relevant for LHC phenomenology.
As a technical benchmark we provide in eqs.~\eqref{eq:MIvalues}
the values of these five integrals, at $\sqrt{s}=120$ GeV and $\cos\theta=0$,
using the $Z$ boson complex mass with the values presented in the next section.
\bea
MI_{32}&=&
(  -   0.1749432674210653458206641571672526693701  \label{eq:MIvalues}\\   
&& -  0.08577133842383776228342363457748528348231  ~i ) \frac{1}{\varepsilon} + \nonumber\\
&&(  - 0.6880611799340122002547431584730743961212 \nonumber\\ 
 &&  - 0.7172116087719097073815199889325839829478 ~i )\nonumber\\
MI_{33}&=&
(   + 0.04590321278523478276320861551168882058745 \nonumber\\
 && -0.001565827650440604122381753305492983831372  ~i ) \frac{1}{\varepsilon^2} + \nonumber\\
&&( -  0.1020302461263331342976597396613145347408 \nonumber\\
 && + 0.02750120560010053399615654731836551427554  ~i ) \frac{1}{\varepsilon} + \nonumber\\
&&( -  0.9440798936974029538680877301057978456673 \nonumber\\ 
  &&-  0.7373834572088279116262037334972425113889  ~i )\nonumber\\
MI_{34}&=&
(   - 0.04296293184352405065819376225216006379338 \nonumber\\   
 &&+0.0009207027182454070378064541717970157008000  ~i ) \frac{1}{\varepsilon^2} + \nonumber\\
&&( + 0.08602650524260701502345707352625809659972 \nonumber\\ 
 && -  0.1041839205419216009927204337308113344692  ~i )  \frac{1}{\varepsilon} + \nonumber\\
&&( +  0.6256624139686838528066374865168826034798 \nonumber\\
 && +  0.2884434585316438395669902051016636313148  ~i ) \nonumber\\ 
MI_{35}&=&
(  -  0.03978405113517657583808686702509598723620 \nonumber\\      
&& +0.0002682033570842286848993717924773930706110  ~i ) \frac{1}{\varepsilon^2} + \nonumber\\
&&(-  0.08594509272571227733490076066555616396611 \nonumber\\   
&& -   0.1071667563486545058940834480592238792738  ~i )  \frac{1}{\varepsilon} + \nonumber\\
&&(+   0.1863097462278093777896412913148906045862 \nonumber\\ 
&& -   0.1309746755875472477456608996344754633763  ~i ) \nonumber\\
MI_{36}&=&   
(  +  0.03722281743928349433413785307317038487389 \nonumber\\ 
&& +0.0002207346628993822499145750213845713169838  ~i )  \frac{1}{\varepsilon^2} + \nonumber\\
&&(+  0.05861962174451141932860447142100263348096 \nonumber\\ 
&& +   0.1128699536667789677655115587719042497769   ~i )  \frac{1}{\varepsilon} + \nonumber\\
&&(-   0.2459055670537880464977311941324743072103 \nonumber\\
&& - 0.004310307792522490155395154024385441988819  ~i )\nonumber
\eea
This approach faces all the problems of the analytical continuation
immediately from the analysis of the structure of the differential equations.
If it is possible to expand the solution in a region around
the boundary condition points,
then the problem becomes of algebraic nature,
i.e. we have to solve for all the coefficients of the various powers
of the expansion variable.
The system of differential equations of ref.~\cite{Bonciani:2016ypc}
can be cast in dlog form and thus contains the complete information
about the singularity structure of the problem,
in particular the position of the branch points relevant
in the discussion of the polylogarithmic functions present in the solution.
This information can be exploited to control the analytic continuation when
we extend the solution to an adjacent region, either non-physical or physical.
We remark that complete control on the analytic continuation has been verified
when letting vary one kinematical variable at the time,
while the others are kept fixed.

\subsection{Numerical results}
\label{sec:numerics}
In this Section, we present the numerical evaluation of our result,
the finite IR-subtracted UV-renormalised hard function ${H}^{(1,1)}$.
For every practical application,
the evaluation in one phase-space point of the correction factor
stemming from the two-loop \oaas virtual corrections
requires a cumbersome procedure.
We consider it more convenient to prepare a numerical grid for
${H}^{(1,1)}$,
covering all the phase-space values relevant for
NC DY in a given fiducial volume and parameterised in terms of
the partonic centre-of-mass energy $\sqrt{s}$ and scattering angle $\cos\theta$.
We notice that the virtual correction,
after the subtraction of all the IR enhanced factors, is a smooth,
slowly varying function of  $\sqrt{s}$ and $\cos\theta$.
We have prepared a grid with a sampling based on the known behaviour
of the NLO-EW distribution,
with special care for the $Z$ resonance region,
where a finer binning is necessary.
\begin{figure}[t]
\begin{center}
\includegraphics[width=7.5cm]{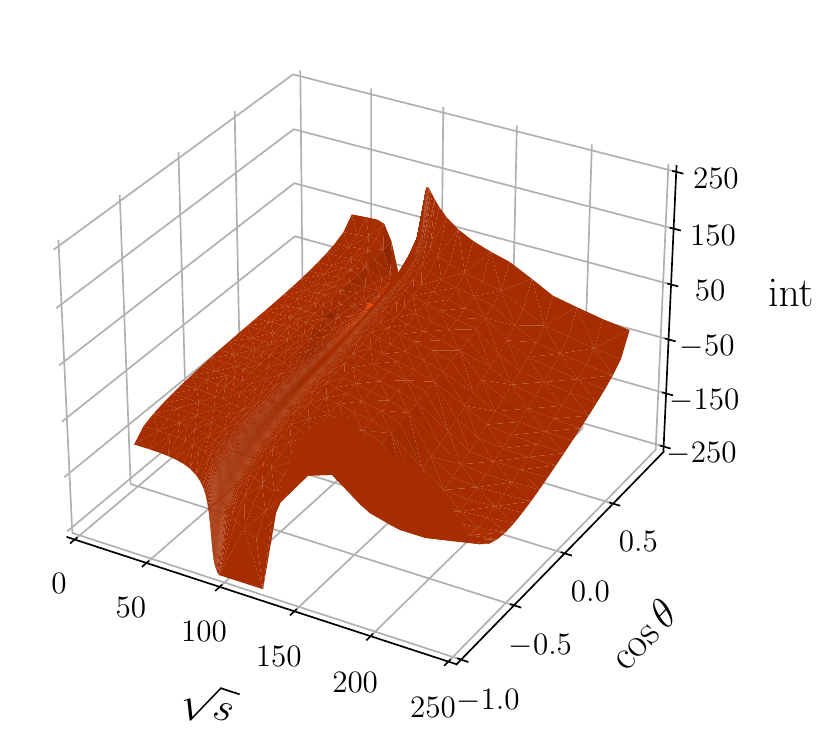}
\includegraphics[width=7.5cm]{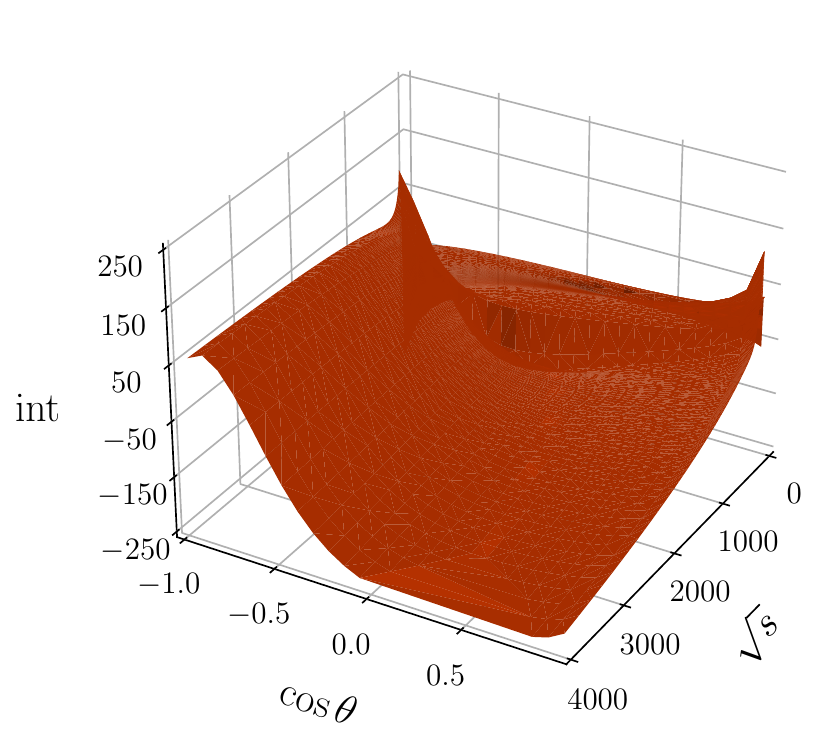}
\end{center}
\vspace{-2ex}
\caption{\label{fig:correction}
 The contributions from all the vertex and box Feynman diagrams
 to the \oaas correction to the finite hard function,
  in two different invariant mass ranges,
  as a function of $\sqrt{s}$ and $\cos\theta$.
  }
\end{figure}
For illustration purposes we consider here a grid with (130x25) points
in $(\sqrt{s},\cos\theta)$,
covering the intervals $s\in [40,13000]$ GeV and $\cos\theta\in [-1,1]$
and present the values in the ancillary file \textsf{gridsth.m}\,.
The extraction of any intermediate value
of ${H}^{(1,1)}$
can be obtained via interpolation of the grid points
with excellent accuracy thanks to the smoothness of the function.
  With the sampling provided as ancillary file,
  we reproduce the exact correction with an accuracy at the $10^{-3}$ level, in the whole phase space.
  This accuracy is sufficient for phenomenological studies, but can be increased whenever needed.

  The preparation, on a single core, of a grid of values for all the 36 two-masses  MIs requires a fixed step, executed only once,
  to move from the boundary conditions fixed in the non-physical euclidean region to the physical region, which can take
   $\sim 20$ minutes. Moving in the physical region is faster, with an average time of $\sim 5$ minutes per point.
  Once this first grid is ready, the evaluation of the corresponding grid for the $H^{(1,1)}$ function requires
  $\sim 3$ minutes per point.
  Once the  $H^{(1,1)}$ grid is ready, its interpolation in any simulation code requires a negligible time.

We use the following parameters:
\begin{center}
 \begin{tabular}{c l l l}
 \hline\hline
  $M_Z$ & 91.1535 GeV & $\Gamma_Z$ & 2.4943 GeV\\
  $M_W$ & 80.358 GeV  & $\Gamma_W$ & 2.084 GeV\\
  $m_H$ & 125.09 GeV  & $m_t$ & 173.07 GeV\\
 \hline\hline
 \end{tabular}
\end{center}
The values of the gauge boson masses and decay widths
      are compatible with those reported in the PDG \cite{ParticleDataGroup:2020ssz},
      extracted in a running-width fitting scheme.
The choice of the complex mass scheme to define the masses
of the unstable $W$ and $Z$ bosons
implies that all the functions which depend on these parameters
become functions of complex variables.
We have exploited the rescaling properties of the GHPLs
to set their argument equal to one, adjusting all the other weights,
which become in general complex\footnote{Cfr. also ref. \cite{Bonciani:2010ms}}.
For the numerical evaluation of all the GHPLs
appearing in the final expressions,
we use {\tt GiNaC} and
{\tt handyG} \cite{Naterop:2019xaf}
in two independent {\tt C++} and {\tt Mathematica} codes
to evaluate each phase-space point.
We also use {\tt HarmonicSums} \cite{Ablinger:2010kw,Ablinger:2014rba}, {\tt PolyLogTools} \cite{Duhr:2019tlz}
  and {\tt LoopTools} \cite{Hahn:1998yk}
  for several cross-checks of the one- and two-loop integrals available in closed analytic form,
  with real and complex masses.
For the numerical evaluation of the 5 MIs of eq.~(\ref{eq:boxmis}),
we use the semi-analytical approach described
in Section \ref{sec:semianalytical}.

We present in Figure \ref{fig:correction}
the numerical grid expressing the UV-renormalised IR-subtracted
two-loop \oaas virtual correction
due to the two-loop vertex and box diagrams (Figure \ref{fig:diagsamples} a,b,f and Figure \ref{fig:diagsamplesCT} a,b,f).
The plots show the correction factor normalised to the Born cross section
and expressed in units $\frac{\alpha}{\pi} \frac{\alpha_s}{\pi}$,
as a function of $\sqrt{s}$ and $\cos\theta$.
We consider the range $-1\leq\cos\theta\leq 1$ and,
for the partonic center-of-mass energy, two different intervals,
namely $50\leq\sqrt{s}\leq 250$ GeV and $50\leq\sqrt{s}\leq 5000$ GeV.

With the zoom in the $50\leq\sqrt{s}\leq 250$ GeV interval,
we can appreciate the non-trivial structure of the corrections
starting from the $Z$ resonance and up to 110 GeV.
The description of the $WW$ and $ZZ$ thresholds is smooth,
thanks to the implementation of the complex-mass scheme.
Above 200 GeV the shape of the correction factor is negative, smooth,
and slowly decreasing;
we do not observe the appearance of additional structures.
The non-trivial $\cos\theta$ dependence is due to parity violating effects,
in largest size stemming from the interference between the two-loop
$AA$ subset of diagrams and the tree-level amplitude with a $Z$-boson exchange.

\section{Conclusions}
\label{sec:conclusions}
In the NLO case, a complete automation of all the steps of
the calculation is in a quite mature stage for both QCD and EW corrections.
In the  NNLO case instead, only parts of the procedure are automatic,
because of various conceptual and computational problems.
In this paper we have presented the details of the complete calculation
of the exact \oaas virtual corrections to the NC DY process,
describing how we have overcome the problems
appearing in the different stages of the procedure
that brings us from the Feynman diagrams
to the numerical evaluation of the squared matrix element.

All the algebraic manipulation processes face the difficulty of dealing with
potentially very large expressions, so that a systematic simplification
work, based on the search for recurring patterns,
is crucial to keep the size of the formulae at a manageable level.
While the reduction to MIs is completely standardised,
the analytical subtraction of the IR divergences
is not mature at the same level,
and we have achieved this successful check only via a dedicated calculation.
The presence of internal massive lines yields a rich analytical structure
of the Feynman integrals;
for the evaluation of the two-loop boxes with two massive lines
we have adopted a semi-analytical approach, valid for arbitrary complex masses.

The final correction factor, defined in the $q_T$-subtraction scheme,
has been cast in a relatively compact form,
which we provide in the ancillary files.
This finite virtual factor can be used in any simulation code,
provided that the appropriate finite correction for the IR subtraction scheme
is applied.
For benchmarking we provide the grid with the numerical evaluation
of the correction factor in the phase-space region relevant at the LHC.

The successful completion of all these steps can help to devise
new strategies for the automation of the calculation of two-loop
EW and mixed QCD-EW radiative corrections,
relevant for processes at the LHC and future lepton colliders.

\section*{Acknowledgments}
We would like to thank L. Buonocore, M. Grazzini, S. Kallweit, C. Savoini and F. Tramontano,
for several interesting discussions during the development of the complete
calculation of the NNLO QCD-EW corrections to NC DY and for a careful reading of the manuscript.
We would like to thank G. Heinrich for help with {\sc pySecDec}
and R. Lee for help with {\sc LiteRed}.

S.D., N.R. and A.V. are supported by the Italian Ministero della Universit\`a e della Ricerca (grant PRIN201719AVICI\_01).
R. B. is partly supported by the italian Ministero della Universit\`a e della Ricerca (MIUR) under grant PRIN 20172LNEEZ.

\appendix
\section{Description of the ancillary files}
\label{app:readme}
In this subsection we present a description of the content of
the ancillary files attached to the paper.
A complete listing of all the symbols used in the analytical expressions
and their correspondence with the physical quantities can be found in the
\textsf{README} file.

In the ancillary {\tt Mathematica} file \textsf{NCDY\_ME11.m},
we present $\langle \M^{(0)} | \M^{(1,1),fin} \rangle$ i.e.
the finite IR subtracted and UV renormalised matrix elements of all the vertex
and box type Feynman diagrams.
As discussed earlier, we present the result in the small lepton mass limit
i.e. dropping all the terms which vanish as $m_l \rightarrow 0$.
In the ancillary file \textsf{uU\_totaldelta\_vertbox.dat},
we provide the numerical evaluation, for up-type quark initiated channel, of
the expressions contained in \textsf{NCDY\_ME11.m},
divided by Born, multiplied by 2, divided by 16 and also subtracted of the logarithms of the lepton mass,
using the numerical value of the input parameters as described in Section
\ref{sec:numerics},
for all the points of the grid specified in \textsf{gridsth.m}.
We also provide the exact result
including the full dependence of the lepton mass
in \textsf{ME11\_partml\_exact.m}
for the following subsets:
all the vertex and box type contributions for the $\gamma \gamma$ subset
and the contributions from the Feynman diagrams
with NLO QED correction to the lepton vertex in the $\gamma Z$ subset.
The explicit expressions
of the two-loop self energy and one-loop self energy with NLO QCD contributions,
are known in the literature (cfr. Refs.~\cite{Dittmaier:2020vra} and Refs.~\cite{Denner:1991kt}).
Hence, we refrain ourselves from providing these results.

The results are written as a combination of GHPLs and of the symbols associated
to the MIs whose solution has been obtained via a semi-analytical method.
Several variables have been introduced to define the weights and arguments of the GHPLs,
which we report here for the sake of definiteness.
Extending the notation of eq.~(\ref{eq:defxl}),
the Landau variables $x_m$ are defined with respect to the mass $m$ as follows,
\begin{equation}
  \frac{(1-x_m)^2}{x_m} = -\frac{s}{m^2}
  \label{eq:defxm}
\end{equation}
where $m \in \{ m_l, \mz, \mw \}$.
The kinematical ratios are defined as
\begin{equation}
  \bar{x}_m = -\frac{s}{m^2}   \,, ~~
 y_m =  -\frac{t}{m^2}  \,, ~~
 z_m = -\frac{u}{m^2}  \,.
\end{equation}
We define other variables $v_m$ and $v_m^{'}$ as follows
\begin{equation}
 -\frac{t}{m^2} = \frac{x_m}{v_m} \frac{(1+v_m)^2}{(1+x_m)^2} \,, \quad
 -\frac{u}{m^2} = \frac{x_m}{v_m^{'}} \frac{(1+v_m^{'})^2}{(1+x_m)^2} \,.
\end{equation}
The arguments of the GHPLs in the expressions have been reduced to 1
(denoted by `\textsf{one}').
The GHPLs are expressed as
\begin{equation}
 G[a\_\_,1] \equiv L[\{a\}, \mathsf{one}] \,.
\end{equation}
The coefficients of $\ep^n$ of the series expansion in $\ep$ of the 5 MIs
of eq.~(\ref{eq:boxmis}) for topology $B_{16}$ with $\mu_V = \mz$ are denoted
by
\begin{equation}
 \mathsf{B16z[I32,n]}, \mathsf{B16z[I33,n]}, \mathsf{B16z[I34,n]},
\mathsf{B16z[I35,n]}, \mathsf{B16z[I36,n]}\,.
\end{equation}
For topology $B_{16p}$ with $\mu_V = \mz$, they are denoted by
\begin{equation}
 \mathsf{B16zp[I32,n]}, \mathsf{B16zp[I33,n]}, \mathsf{B16zp[I34,n]},
\mathsf{B16zp[I35,n]}, \mathsf{B16zp[I36,n]}\,.
\end{equation}
For topology $B_{16p}$ with $\mu_V = \mw$, they are denoted by
\begin{equation}
 \mathsf{B16wp[I32,n]}, \mathsf{B16wp[I33,n]}, \mathsf{B16wp[I34,n]},
\mathsf{B16wp[I35,n]}, \mathsf{B16wp[I36,n]}\,.
\end{equation}
The result also contains the one-loop box integral with two massive lines (Figure 5-${\cal T}_6$ of ref.~\cite{Bonciani:2016ypc}) symbolically.
We denote the $\ep^n$ coefficient of this one-loop box integral as
\begin{align}
 \mathsf{A16z[I6,n]}, \mathsf{A16zp[I6,n]}, \mathsf{A16wp[I6,n]}
\end{align}
where the one-loop topologies are defined by
\begin{align}
  {\rm A}_{16} &: \{ \cD_1-\mu_V^2, \cD_{1;1}, \cD_{1;12}-\mu_V^2, \cD_{1;3} \}
              \nonumber\\
  {\rm A}_{16p} &: \{ \cD_1-\mu_V^2, \cD_{1;2}, \cD_{1;12}-\mu_V^2, \cD_{1;3} \}\,.
\end{align}
The representation of the coupling constants and charges in the expressions
is detailed below.
$C_{v,f}$ and $C_{a,f}$ are the charges of the coupling of the Z boson to a
fermion $f$.
They are given by
\begin{equation}
 C_{v,f} = \left( \frac{I_W^{(f)}}{2} - Q_f \, \sin \theta_W  \right)\,, ~~~
 C_{a,f} = \left( \frac{I_W^{(f)}}{2}  \right)\,.
\end{equation}
$I_W^{(f)}$ and $Q_f$ are the third component of the weak isospin and the
electric charge in units
of the positron charge, respectively. $\theta_W$ is the weak mixing angle.
The following symbols and abbreviations are used throughout the files.
\begin{align}
&
\mathsf{asr} = \frac{\alpha_s}{4\pi}; ~~
\mathsf{aem} = \frac{\alpha}{4\pi}; ~~
\mathsf{borncf} = N_C; ~~
\mathsf{Cf} = C_F; ~~
\mathsf{gA4} = e^4; ~~
\nonumber\\&
\mathsf{Qu} = Q_u; ~~~
\mathsf{Qd} = Q_d; ~~~
\mathsf{Ql} = Q_l; ~~~
\nonumber\\&
\mathsf{gVu} = C_{v,u}; ~~
\mathsf{gAu} = C_{a,u}; ~~
\mathsf{gVd} = C_{v,d}; ~~
\mathsf{gAd} = C_{a,d}; ~~
\nonumber\\&
\mathsf{gVl} = C_{v,l}; ~~
\mathsf{gAl} = C_{a,l}; ~~
\mathsf{gVn} = C_{v,n}; ~~
\mathsf{gAn} = C_{a,n}; ~~
\end{align}
\noindent
\textsf{gVuC}, \textsf{gAuC}, \textsf{gVlC}, \textsf{gAlC} are complex
conjugate of \textsf{gVu}, \textsf{gAu}, \textsf{gVl}, \textsf{gAl},
respectively.
The following constants appear in the expressions
\begin{align}
\mathsf{pi} = \pi; ~~
\mathsf{IPi} = i \pi; ~~
\mathsf{z2} = \zeta_2; ~~
\mathsf{z3} = \zeta_3;
\mathsf{c}  = \frac{1}{2} + i \frac{\sqrt{3}}{2}; ~
\mathsf{cb} = \frac{1}{2} - i \frac{\sqrt{3}}{2};
\end{align}
The presentation of the variables appearing in the expression is as follows:
\begin{align}
&
\mathsf{xrl} = \bar{x}_{m_l}; ~~~
\mathsf{xl} = x_{m_l}; ~~~
\mathsf{yl} = y_{m_l}; ~~~
\mathsf{zl} = z_{m_l}; ~~~
\nonumber\\&
\mathsf{xrZ} = \bar{x}_{\mz}; ~~~
\mathsf{xZ} = x_{\mz}; ~~~
\mathsf{yZ} = y_{\mz}; ~~~
\mathsf{zZ} = z_{\mz}; ~~~
\mathsf{vZ} = v_{\mz}; ~~~
\mathsf{vpZ} = v^{'}_{\mz}; ~~~
\nonumber\\&
\mathsf{xrW} = \bar{x}_{\mw}; ~~
\mathsf{xW} = x_{\mw}; ~~
\mathsf{yW} = y_{\mw}; ~~
\mathsf{zW} = z_{\mw}; ~~
\mathsf{vW} = v_{\mw}; ~~
\mathsf{vpW} = v^{'}_{\mw};
\nonumber\\&
\mathsf{isqvZ} = i \sqrt{v_{\mz}}; ~~~
\mathsf{isqvpZ} = i \sqrt{v^{'}_{\mz}}; ~~~
\mathsf{isqvW} = i \sqrt{v_{\mw}}; ~~
\mathsf{isqvpW} = i \sqrt{v^{'}_{\mw}};
\nonumber\\&
\mathsf{cBpropml} = \frac{m_l^2}{s-\cmz^2}; ~
\mathsf{cBpropmZ} = \frac{\mz^2}{s-\cmz^2}; ~
\mathsf{cBpropmW} = \frac{\mw^2}{s-\cmz^2};
\end{align}
$\cmz$ is the complex conjugate of $\mz$.

\bibliography{long}
\bibliographystyle{JHEP}

\end{document}